\documentclass[aps,prl,twocolumn,groupedaddress,showpacs,amsmath]{revtex4-1}

\usepackage{xfrac}

\begin{document}


\title{Entanglement and Sources of Magnetic Anisotropy in Radical Pair-Based Avian Magnetoreceptors}


\author{Hannah J. Hogben}
\affiliation{Department of Chemistry, University of Oxford, Physical \&\ Theoretical Chemistry Laboratory, Oxford, OX1 3QZ, UK.}
\author{Till Biskup}
\affiliation{Department of Chemistry, University of Oxford, Physical \&\ Theoretical Chemistry Laboratory, Oxford, OX1 3QZ, UK.}
\author{P. J. Hore}
\email[]{peter.hore@chem.ox.ac.uk}
\affiliation{Department of Chemistry, University of Oxford, Physical \&\ Theoretical Chemistry Laboratory, Oxford, OX1 3QZ, UK.}


\date{\today}

\begin{abstract}
One of the principal models of magnetic sensing in migratory birds rests on the quantum spin-dynamics of transient radical pairs created photochemically in ocular cryptochrome proteins. We consider here the role of electron spin entanglement and coherence in determining the sensitivity of a radical pair-based geomagnetic compass and the origins of the directional response. It emerges that the anisotropy of radical pairs formed from spin-polarized molecular triplets could form the basis of a more sensitive compass sensor than one founded on the conventional hyperfine-anisotropy model. This property offers new and more flexible opportunities for the design of biologically inspired magnetic compass sensors.
\end{abstract}

\pacs{03.67.Bg, 82.30.Cf, 87.50.C-}

\maketitle


The biophysics and biochemistry that allow birds to sense the direction of the geomagnetic field (25-65 $\mu$T) are for the most part obscure. One of the two currently popular hypotheses (the other involves biogenic iron-oxide nanostructures \cite{wink-jrsif-7-S273}) is founded on magnetically sensitive photochemical reactions in the retina \cite{schu-zpc-111-1}. It is thought that photo-induced radical pairs in cryptochrome, a blue-light photoreceptor protein, may constitute the primary magnetic sensor \cite{ritz-bj-78-707,maed-pnas-109-4774} and a variety of supporting evidence has accumulated over the last few years (reviewed in \cite{rodg-pnas-106-353,lied-jrsif-7-S147,ritz-pchem-3-262,mour-conb-22-343}). If this mechanism proves to be correct, it will incontrovertibly come under the umbrella of `quantum biology' \cite{ball-n-474-242}, as an instance of Nature using fundamentally quantum behaviour -- in this case the coherent spin dynamics of radical pairs -- to achieve something that would be essentially impossible by means of more conventional chemistry. For this reason, the avian magnetic compass has attracted the attention of quantum information theorists and others wishing to understand the role played by spin-entanglement and to determine whether the techniques of quantum control could shed light on this intriguing sensory mechanism \cite{cai-prl-104-220502,gaug-prl-106-040503,cai-pra-85-022315,cai-pra-85-040304}. 

A fundamental property of radical pairs that allows sensitivity to magnetic interactions orders of magnitude smaller than $k_{\rm B}T$ is that their chemical transformations conserve electron spin. Radical pairs are therefore created with the same spin-multiplicity (singlet or triplet) as their precursors. Owing to electron-nuclear hyperfine (HF) interactions, neither singlets nor triplets are, in general, eigenstates of the spin Hamiltonian. Consequently, the radical pair starts out in a non-stationary superposition which evolves coherently at frequencies determined by the HF interactions and also, crucially for a magnetic sensor, by the electronic Zeeman interactions with an external magnetic field \cite{rodg-pnas-106-353}. Spin decoherence and spin relaxation can be slow enough to allow even an Earth-strength magnetic field to modulate the spin dynamics and hence alter the yields of the products formed by spin-selective reactions. The anisotropy of the HF interactions leads to anisotropic reaction yields and hence, in principle, a magnetic direction sensor \cite{cint-cp-294-385,maed-n-453-387}.

The singlet state -- the initial state of the radical pairs formed photochemically in cryptochromes \cite{maed-pnas-109-4774,webe-jpcb-114-14745} -- is entangled:
\begin{eqnarray} \label{eq:singletstate}
|{\rm S}\rangle\langle{\rm S}| &=& \tfrac{1}{2}|\alpha_1\beta_2\rangle\langle\alpha_1\beta_2|+\tfrac{1}{2}|\beta_1\alpha_2\rangle\langle\beta_1\alpha_2|\\\nonumber
&&-\tfrac{1}{2}|\alpha_1\beta_2\rangle\langle\beta_1\alpha_2|-\tfrac{1}{2}|\beta_1\alpha_2\rangle\langle\alpha_1\beta_2|
\end{eqnarray}
($\alpha$ and $\beta$ are the $m_S=\pm\frac{1}{2}$ spin states of the two unpaired electrons).  But other initial states are also known to result in magnetically sensitive chemistry \cite{stei-cr-89-51}: do they too need to be entangled or is it sufficient if they are `merely' coherent? Or is neither entanglement nor coherence necessary for a magnetic compass?

Questions such as these have been addressed in two recent papers. Briegel and his group noted that randomly generated separable (i.e. not entangled) initial states could result in reaction product yields more anisotropic than those produced from an initial singlet state under the same conditions \cite{cai-prl-104-220502}. The other study, by Benjamin and colleagues, reached similar conclusions by analysing model radical pair systems, finding significant product yield anisotropies for the separable initial state \cite{gaug-prl-106-040503}
\begin{eqnarray} \label{eq:separablestate}
\tfrac{1}{2}|{\rm S}\rangle\langle{\rm S}| + \tfrac{1}{2}|{\rm T_0}\rangle\langle{\rm T_0}| &=& \tfrac{1}{2}|\alpha_1\beta_2\rangle\langle\alpha_1\beta_2|\\\nonumber
&& + \tfrac{1}{2}|\beta_1\alpha_2\rangle\langle\beta_1\alpha_2|
\end{eqnarray}
in which ${\rm T}_0$ is the $m_S=0$ triplet spin state.

Here we examine the role of initial entanglement and attempt to clarify the various sources of magnetic anisotropy that might form the basis of a radical pair compass sensor in birds.

\paragraph{Initial radical pair states.}

We start by identifying chemically feasible initial electron spin states. Geminate radical pairs are normally formed by spin-conserving chemical reactions so that at the moment of their creation they are either pure singlet, described by the initial electron spin density matrix $\hat{\rho}_0=\hat{\rho}_0({\rm S})=|{\rm S}\rangle\langle{\rm S}|$, or pure triplet $\hat{\rho}_0=\hat{\rho}_0({\rm T})=\frac{1}{3}\left(\hat{\openone}-|{\rm S}\rangle\langle{\rm S}|\right)$. Occasionally, singlet and triplet formation channels operate in parallel \cite{maed-cc-47-6563}, in which case $\hat{\rho}_0$ is a weighted sum of $\hat{\rho}_0({\rm S})$ and $\hat{\rho}_0({\rm T})$, i.e. of $|{\rm S}\rangle\langle{\rm S}|$ and $\hat{\openone}$:
\begin{eqnarray}\label{eq:mixedstart}
\hat{\rho}_0 &=& \mu\hat{\rho}_0({\rm S})+(1-\mu)\hat{\rho}_0({\rm T})\\\nonumber
&=& \tfrac{1}{3}(4\mu-1)|{\rm S}\rangle\langle{\rm S}|+\tfrac{1}{3}(1-\mu)\hat{\openone}  
\end{eqnarray}
Eq.~\eqref{eq:mixedstart} is also appropriate for `F-pairs' \cite{stei-cr-89-51} formed from radicals with uncorrelated spins (i.e. $\mu=\frac{1}{4}$). The operators $|{\rm S}\rangle\langle{\rm S}|$ and $\hat{\openone}$ and their linear combinations are invariant to rotations in the electron spin-space, meaning that all states that can be written in the form of Eq.~\eqref{eq:mixedstart} are isotropic. Any $\hat{\rho}_0$ that cannot be so written is necessarily anisotropic. 

Significantly different initial states can occur when the radical pair comes from a molecular triplet precursor formed by intersystem crossing (ISC). This route is common in photochemical reactions of the general type: 
\begin{eqnarray} \label{eq:reactionscheme}
\text{AB}\xrightarrow{\;h\nu\;}\,^{\rm S}[{\rm AB}]^\ast\xrightarrow{\;\text{ISC}\;}\,^{\rm T}[{\rm AB}]^\ast\xrightarrow{\;\text{reaction}\;}{^{\rm T}[{\rm A^\bullet\;B^\bullet}]}
\end{eqnarray}
in which the final step that creates the triplet radical pair could be homolysis (as shown) or inter- or intramolecular electron transfer, hydrogen atom transfer, etc. The formation of $^{\rm T}[{\rm AB}]^\ast$ from $^{\rm S}[{\rm AB}]^\ast$ requires the creation of spin angular momentum at the expense of orbital angular momentum. This process is mediated by spin-orbit coupling and is anisotropic in the molecular frame \cite{groo-mp-12-259}. That is, the three triplet sub-levels of $^{\rm T}[{\rm AB}]^\ast$ are differentially populated leading to a spin polarization in the molecular frame that is passed to the radical pair on its formation. In an appropriately chosen molecular axis system, the initial state of the radical pair may be written:
\begin{eqnarray} \label{eq:molframe}
\hat{\rho}_0 &=& \sum_{q=x,y,z}p_q |{\rm T}_q\rangle\langle{\rm T}_q|
\end{eqnarray}
Anisotropic ISC is known to be responsible for a variety of spin-chemical and spin-polarization phenomena \cite{stei-cr-89-51,atki-mp-27-1633,kats-mp-100-1245,koth-jpcb-114-14755}. 

Aside from linear combinations of Eqs~\eqref{eq:mixedstart} and \eqref{eq:molframe}, there are no other commonly occurring initial conditions for radical pairs subject to weak magnetic fields.

\paragraph{Minimal radical pair model.}

Insights into the spin dynamics of the various initial states just identified can be obtained from a minimal model \cite{timm-mp-95-71} comprising two electron spins one of which is coupled to a spin-\sfrac{1}{2} nucleus (e.g. $^1$H). The HF interaction is either isotropic or axially anisotropic according to the value of a dimensionless parameter, $\alpha$ \cite{cint-cp-294-385}. Two cases are considered specifically: $\alpha=0$ (isotropic) and $\alpha=-1$ (the anisotropic interaction that results in the largest reaction yield anisotropy for this 3-spin system \cite{cai-pra-85-040304}). To account for the chemical reactivity of the radical pair, we adopt the `exponential model' \cite{timm-mp-95-71} in which singlet and triplet states react spin-selectively with the same first-order rate constant, $k$, to form distinct products. The quantum yields of these competing reactions are calculated using standard methods \cite{cint-cp-294-385,timm-mp-95-71} (outlined in the Appendix). The two quantities of interest are $\Phi_{\rm S}$, the fractional yield of the product formed via the singlet pathway, referred to here as the `reaction yield', and  $\Delta\Phi_{\rm S}$, the magnitude of its anisotropy: $\Delta\Phi_{\rm S}=\text{max}\left\{\Phi_{\rm S}\right\}-\text{min}\left\{\Phi_{\rm S}\right\}$. The variation of $\Phi_{\rm S}$ with the orientation of the radical pair in a 50 $\mu$T magnetic field is the basis of the compass sensor.

To begin, we choose the isotropic initial condition in Eq.~\eqref{eq:mixedstart} together with an anisotropic HF interaction ($\alpha=-1$). In the not unrealistic limit, $|a|\gg\omega\gg k$ \cite{cint-cp-294-385,timm-cpl-334-387}:
\begin{eqnarray} \label{eq:isotropic}
\Phi_{\rm S} &=& \tfrac{1}{4}+\tfrac{1}{12}(4\mu-1)\cos^2\theta; \quad \Delta\Phi_{\rm S}=\tfrac{1}{12}|4\mu-1|
\end{eqnarray}
where $a$ is the isotropic HF coupling constant, $\omega$ is the strength of the magnetic field, and $\theta$ is the angle between the symmetry axis of the HF tensor and the magnetic field vector. $\Phi_{\rm S}$ is anisotropic, and therefore potentially suitable as a magnetic compass, except when the initial state is a statistical (\sfrac{1}{4} : \sfrac{3}{4}) mixture of singlet and triplet ($\mu=\frac{1}{4}$). The maximum anisotropy   ($\Delta\Phi_{\rm S}=\frac{1}{4}$) occurs when the initial state is pure singlet ($\mu=1$); for a pure triplet initial state ($\mu=0$), $\Delta\Phi_{\rm S}$ is smaller by a factor of three. These results were verified by exact numerical simulations (see Appendix).

To quantify the entanglement of the various initial electron spin states considered here, we use the `concurrence' $C(\hat{\rho}_0)$ proposed by Wootters \cite{woot-prl-80-2245} for a two-qubit density operator. For the initial condition in Eq.~\eqref{eq:mixedstart}, $C(\hat{\rho}_0)$ is $2\mu-1$ when $\mu>\frac{1}{2}$ and zero when $\mu\le\frac{1}{2}$ (see Appendix). Thus, a singlet--triplet mixture must contain more than 50\% singlet for the initial state to be entangled. The pure triplet state ($\mu=0$) is not entangled, but as we have just seen it gives rise to a significantly anisotropic reaction yield.

We now turn to a different initial condition, a linear combination of Eq.~\eqref{eq:mixedstart} (with $\mu=0$) and Eq.~\eqref{eq:molframe} (with $p_x=p_y=0$; $p_z=1$):
\begin{eqnarray} \label{eq:lincomb}
\hat{\rho}_0 &=& \eta|{\rm S}\rangle\langle{\rm S}|+(1-\eta)|{\rm T}_z\rangle\langle{\rm T}_z|
\end{eqnarray}
i.e. an anisotropic mixed singlet-triplet initial state in which the triplet component is 100\% polarized along the molecular $z$-axis. In the same limit as before ($|a|\gg\omega\gg k$), but now for an \emph{isotropic} HF interaction:
\begin{eqnarray} \label{eq:anisotropic}
\Phi_{\rm S} &=& \tfrac{3}{8}-\tfrac{1}{4}(1-\eta)\sin^2\theta; \quad \Delta\Phi_{\rm S}=\tfrac{1}{4}(1-\eta)
\end{eqnarray}
where $\theta$ is now the angle between the triplet polarization axis ($z$) and the magnetic field vector. The anisotropy is maximised when $\eta=0$ (pure $|{\rm T}_z\rangle$ triplet, $\Delta\Phi_{\rm S}=\frac{1}{4}$) and is at a minimum when $\eta=1$ (pure singlet, $\Delta\Phi_{\rm S}=0$). Once again, these expressions were confirmed by numerical simulations (see Appendix). We note that Eqs~\eqref{eq:isotropic} and \eqref{eq:anisotropic} predict identical maximum directional responses.

The reaction yield is isotropic when $\eta=1$ because then both the initial state $|{\rm S}\rangle\langle{\rm S}|$ and the spin-Hamiltonian are isotropic. The angle-dependence in Eq.~\eqref{eq:anisotropic} clearly arises because the spin dynamics depend on the direction of the magnetic field with respect to the quantization ($z$) axis of the initial  $|{\rm T}_z\rangle$ state \cite{kats-njp-12-085016}. The concurrence of the density operator in Eq.~\eqref{eq:lincomb} is $2\eta-1$ when $\eta\ge\frac{1}{2}$ and $1-2\eta$ when $\eta\le\frac{1}{2}$. Pure singlet and pure $|{\rm T}_z\rangle$ triplet thus have the same degree of entanglement but lead to very different $\Delta\Phi_{\rm S}$. 

Hitherto we have taken the reaction rates of the singlet and triplet states ($k_{\rm S}$ and $k_{\rm T}$) to be identical. Once this restriction is lifted, it is even possible to have magnetic field effects when the initial state is a statistical mixture of singlet and triplet: $\hat{\rho}_0=\frac{1}{4}\hat{\rho}_0({\rm S})+\frac{3}{4}\hat{\rho}_0({\rm T})=\frac{1}{4}\hat{\openone}$. To illustrate this point, simulations for the minimal radical pair with an anisotropic HF coupling are included in the Appendix. $\Delta\Phi_{\rm S}$ is non-zero except when $k_{\rm S}=k_{\rm T}$. That is, a radical pair can exhibit magnetic compass properties even when its initial electron spin state is neither entangled nor coherent. In this case the coherence arises during the spin evolution as a result of the differential reactivity of the singlet and triplet states.

\paragraph{Relation between compass properties and entanglement.}

A complex picture emerges from these simple considerations. Entangled initial states can give small or zero reaction yield anisotropy. Non-entangled initial states can lead to appreciable anisotropy. With two sources of anisotropic reaction yields -- the initial state and the HF interactions -- it is tricky to assess whether entanglement, or coherence in a given basis, is essential for magnetic compass action. For example, replacing $\hat{\rho}_0=|{\rm S}\rangle\langle{\rm S}|$ (Eq.~\eqref{eq:singletstate}) by $\hat{\rho}_0=\frac{1}{2}|{\rm S}\rangle\langle{\rm S}|+\frac{1}{2}|{\rm T}_0\rangle\langle{\rm T}_0|$ (Eq.~\eqref{eq:separablestate}) not only removes the initial entanglement, and the coherence in the $\left\{|\alpha_1\beta_2\rangle,|\beta_1\alpha_2\rangle\right\}$ basis, it also introduces anisotropy that was not present in $|{\rm S}\rangle\langle{\rm S}|$. Similarly, most randomly chosen initial states are anisotropic and some will give a larger $\Delta\Phi_{\rm S}$ than does $|{\rm S}\rangle\langle{\rm S}|$ under identical conditions. In short, it appears that initial entanglement is not a particularly helpful concept when assessing the sensitivity of a radical pair compass; nor is it straightforwardly illuminating to consider the behaviour of artificial initial states.

\paragraph{A radical pair compass based on initial-state anisotropy.}

The above considerations suggest an alternative compass design in which the directionality comes from the initial condition rather than the HF interactions. In the minimal model, the initial state that gives the largest reaction yield anisotropy is $\hat{\rho}_0=|{\rm T}_q\rangle\langle{\rm T}_q|$ where $q=x,y,z$ (see Appendix). We therefore compare $|{\rm T}_q\rangle\langle{\rm T}_q|$ with $|{\rm S}\rangle\langle{\rm S}|$ using exact numerical simulations (see Appendix). The possibility that spin-polarized triplet radical pairs might offer some advantage over singlets has been noted before but without realistic suggestions for the chemical origin of such initial states \cite{kats-njp-12-085016}.

\begin{figure}
\includegraphics[width=3in]{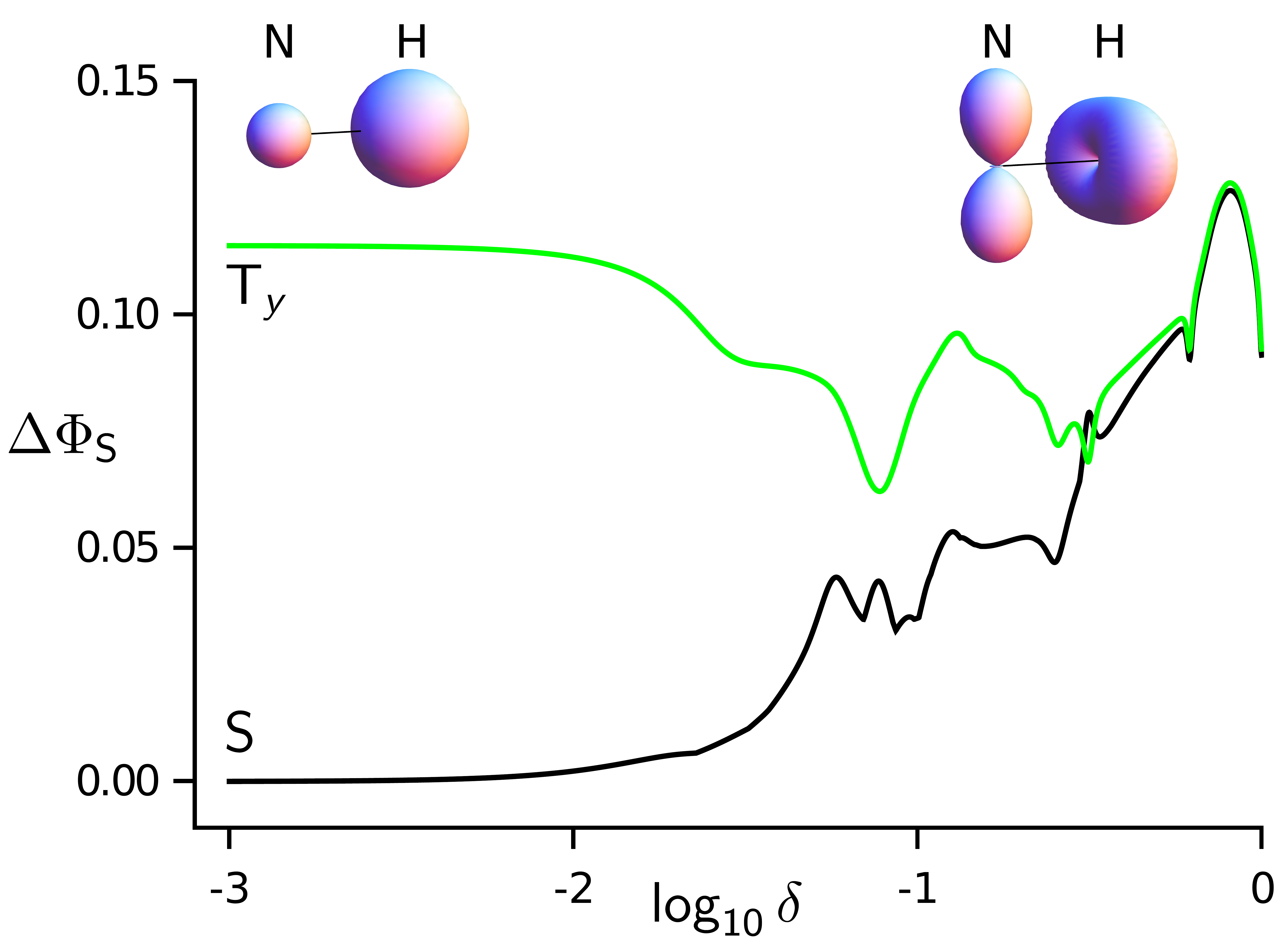}%
\caption{\label{fig:rp11} Reaction yield anisotropy, $\Delta\Phi_{\rm S}$, calculated (see Appendix) for a radical pair in which one radical contains a $^1$H nucleus (spin-\sfrac{1}{2}) and a $^{14}$N nucleus (spin-1). $k=10^6\;\rm s^{-1}$ and $\omega = 50\;\rm\mu T$. The HF coupling parameters (in mT) are: $a_{\rm H} = -0.8$; $T_{{\rm H},xx}  = 0.8\;\delta$; $T_{{\rm H},yy} = -0.6\;\delta$; $T_{{\rm H},zz} = -0.2\;\delta$; $a_{\rm N} = 0.4$; $T_{{\rm N},xx} = -0.5\;\delta$; $T_{{\rm N},yy} = -0.5\;\delta$; $T_{{\rm N},zz} = 1.0\;\delta$. $\hat{\rho}_0=|{\rm S}\rangle\langle{\rm S}|$ (black) and $\hat{\rho}_0=|{\rm T}_y\rangle\langle{\rm T}_y|$ (green). Also shown are representations of the hyperfine tensors for $\delta=0$ (left) and $\delta=1$ (right).}
\end{figure}

Figure~\ref{fig:rp11} shows the reaction yield anisotropy of a radical pair inspired by the flavin adenine dinucleotide radical, FADH$^\bullet$, formed photochemically in cryptochromes \cite{lang-jacs-131-14274}. One radical contains $^1$H and $^{14}$N nuclei with isotropic HF couplings approximately equal to those of the proton and nitrogen (H5 and N5, see appendix) in the central ring of the tricyclic isoalloxazine ring system of FADH$^\bullet$ (these being the two largest HF interactions in FADH$^\bullet$ \cite{webe-jacs-123-3790}). The anisotropic components of the two interactions were also modelled on FADH$^\bullet$, but with a uniform scaling by a factor of $\delta$, in the range $0.001-1.0$. For the smaller values of $\delta$, the spin-Hamiltonian is essentially isotropic. When the initial state $\hat{\rho}_0$ is a 100\% spin-polarized triplet, $\Delta\Phi_{\rm S}$ has significant magnitude for all values of $\delta$. In contrast, when $\hat{\rho}_0=|{\rm S}\rangle\langle{\rm S}|$, $\Delta\Phi_{\rm S}$ is essentially zero until the HF tensors become significantly anisotropic ($\delta\approx0.1$). By the time the HF anisotropy is comparable to that in FADH$^\bullet$ (i.e. $\delta\approx1.0$), both initial states give very similar directional responses to the 50 $\mu$T applied magnetic field. This suggests that a spin-polarized triplet geminate radical pair with isotropic HF interactions could operate as a compass sensor just as well as an initial singlet state with anisotropic HF interactions.

\begin{figure}
\includegraphics[width=3.375in]{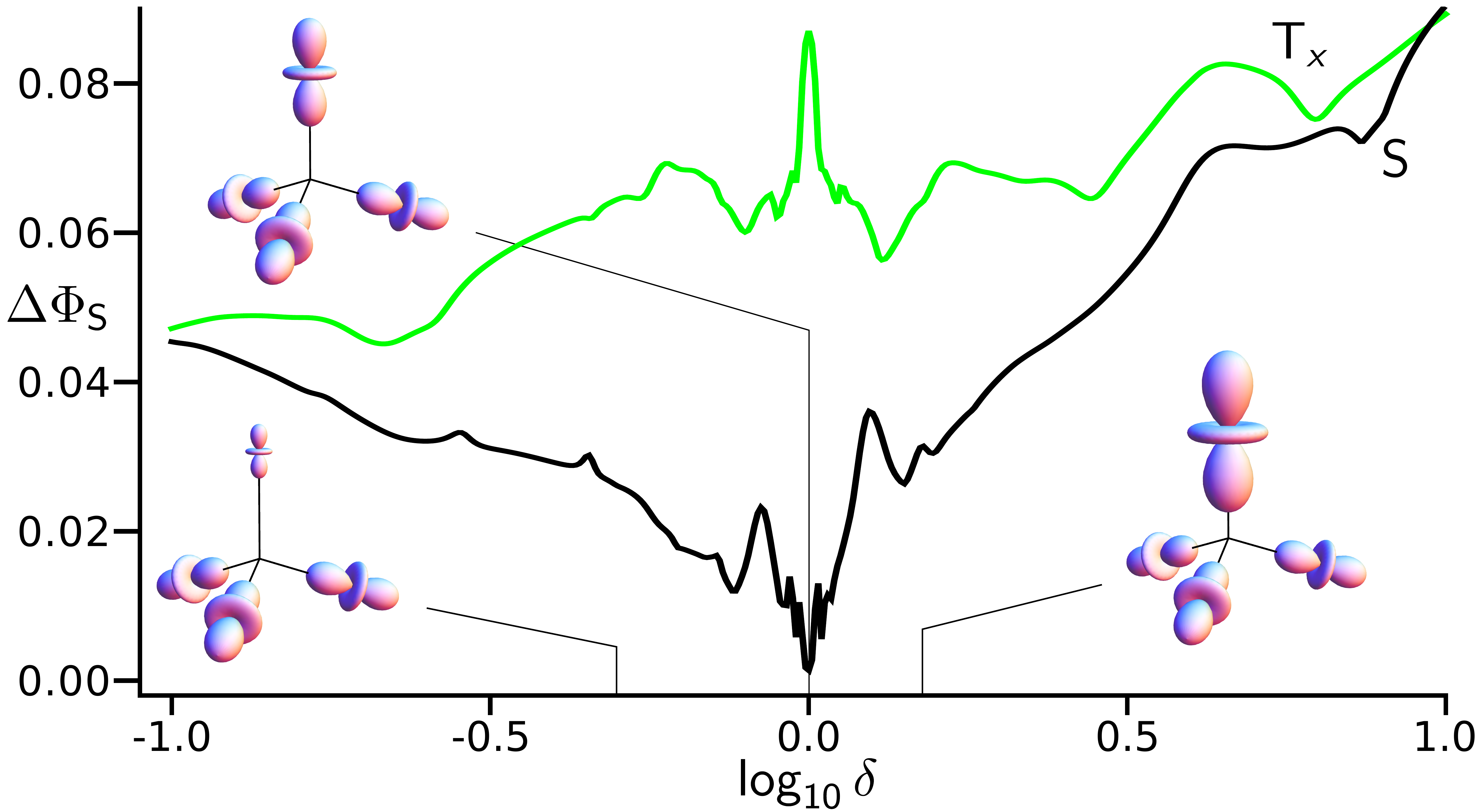}%
\caption{\label{fig:rp40} Reaction yield anisotropy, $\Delta\Phi_{\rm S}$, calculated (see Appendix) for a radical pair in which one radical contains four $^1$H nuclei, all of which have axially anisotropic HF interactions with $a = 0$. The symmetry axes of the four HF tensors are directed towards the vertices of a tetrahedron. Three of the tensors have principal values: $T_{11} = T_{22} = -1.0$, $T_{33} = 2.0\rm\; mT$. The fourth is identical apart from a uniform scaling of the principal values by a factor $\delta$. $k=10^6\;\rm s^{-1}$ and $\omega = 50\rm\;\mu T$. $\hat{\rho}_0=|{\rm S}\rangle\langle{\rm S}|$ (black) and $\hat{\rho}_0=|{\rm T}_x\rangle\langle{\rm T}_x|$ (green). Also shown are representations of the hyperfine tensors for $\delta=0.5$, $\delta=1.0$, and $\delta=1.5$.}
\end{figure}

Indeed, there are circumstances in which, other things being equal, the anisotropy of the initial state might offer a more sensitive compass than one based on HF anisotropy. Biologically plausible radical pairs are likely to have many magnetic nuclei (mostly $^1$H and $^{14}$N) with differently aligned HF tensors. Simulations suggest that the directional information potentially available from individual HF tensors tends to be scrambled in a multi\-nuclear radical pair, resulting in a greatly reduced $\Delta\Phi_{\rm S}$ (see Appendix). A simple illustration of this effect is given in Fig.~\ref{fig:rp40} which shows simulations of the reaction yield anisotropy for a spin system in which one of the radicals contains four spin-\sfrac{1}{2} nuclei with tetrahedrally disposed axial HF tensors. When all four tensors are identical ($\delta=1$), the reaction yield anisotropy for $\hat{\rho}_0=|{\rm S}\rangle\langle{\rm S}|$ vanishes, by symmetry. However, when the symmetry is reduced to $C_{\rm 3v}$, by scaling the principal components of one of the HF tensors by a factor $\delta$, the value of $\Delta\Phi_{\rm S}$ increases but does not approach that afforded by $\hat{\rho}_0=|{\rm T}_x\rangle\langle{\rm T}_x|$ until $|\log_{10}\delta|$ reaches ca. $1.0$. Thus it appears that the compass properties of a radical pair with many mutually cancelling HF interactions could be `rescued' by having a triplet, rather than a singlet, initial condition, provided the triplet is spin-polarized by anisotropic intersystem crossing.

\paragraph{Discussion.}

Having identified the initial spin-states in which radical pairs may be formed by chemical reaction, we revisited earlier attempts to determine the importance of entanglement and coherence as determinants of the anisotropic responses of radical pair magnetoreceptors. It appears that the use of artificial initial spin-states for this purpose is somewhat confounded by their intrinsic anisotropy, the effects of which may dominate the anisotropy conferred by the HF interactions. From these considerations it emerges that the anisotropy of radical pairs formed from spin-polarized molecular triplets could form the basis for a magnetic compass that is more sensitive than one based on the conventional HF-anisotropy model \cite{schu-zpc-111-1} in particular when the HF couplings are not strongly anisotropic or when the individual effects of multiple HF anisotropies tend to counteract one another.

Would a triplet radical pair compass be compatible with cryptochrome as the primary magnetoreceptor?  In the cryptochromes investigated hitherto (bacterial \cite{bisk-acie-50-12647}, plant \cite{maed-pnas-109-4774} and frog \cite{webe-jpcb-114-14745}), flavin-tryptophan radical pairs are formed as \emph{singlets}. However, avian cryptochromes may behave differently, and there are precedents for triplet radical pairs in other flavoproteins \cite{eise-jacs-130-13544,tham-jacs-132-15542}. Superficially, it appears that flavins may be suitable for an initial triplet-state compass: intersystem crossing in both flavin mononucleotide and riboflavin at near-neutral pH results in fractional populations of the zero-field triplet sub-levels of $p_x=\frac{1}{3}, p_y=\frac{2}{3}, p_z=0$ \cite{kowa-jacs-126-11393}. Within the minimal model discussed above, this would lead to a high reaction yield anisotropy, two-thirds that of the maximum possible (see Appendix). 

The use of spin-polarized triplets should open new channels for the design of bio-inspired molecular devices for sensing the direction of weak magnetic fields.

\begin{acknowledgments}
We thank DARPA (QuBE: N66001-10-1-4061) and the EPSRC for financial support.
\end{acknowledgments}

\appendix
\section{Appendix}

\paragraph{Basis states.}

The spin dynamics of radical pairs may usefully be described in terms of two distinct sets of basis states. In both, singlet and triplet states are eigenstates of the total electron spin operator, $\hat{S}$:
\begin{eqnarray} \label{eq:appdx1}
\langle{\rm S}|\hat{S}|{\rm S}\rangle &=& 0\\\nonumber
\langle{\rm T}_i|\hat{S}|{\rm T}_i\rangle &=& \sqrt{2} \quad (i=0,\pm 1\quad\text{or}\quad x,y,z)
\end{eqnarray}
The triplet basis states are either the eigenstates of $\hat{S}_z$, the component of $\hat{S}$ along the $z$-axis:
\begin{eqnarray} \label{eq:appdx2}
\langle{\rm T}_m|\hat{S}_z|{\rm T}_m\rangle &=& m \quad (m=0,\pm 1)
\end{eqnarray}
or are defined in terms of the three cartesian components of $\hat{S}$:
\begin{eqnarray} \label{eq:appdx3}
\langle{\rm T}_q|\hat{S}_q|{\rm T}_q\rangle &=& 0 \quad (q=x,y,z)\\\nonumber
\langle{\rm T}_x|\hat{S}_y|{\rm T}_z\rangle &=& {\rm i} \quad (\text{and cyclic permutations of } x,y,z)
\end{eqnarray}
The relations between the two are:
\begin{eqnarray} \label{eq:appdx4}
|{\rm T}_x\rangle &=& \tfrac{1}{\sqrt{2}}|{\rm T}_{-1}\rangle-\tfrac{1}{\sqrt{2}}|{\rm T}_{+1}\rangle \\\nonumber
|{\rm T}_y\rangle &=& \tfrac{\rm i}{\sqrt{2}}|{\rm T}_{-1}\rangle+\tfrac{\rm i}{\sqrt{2}}|{\rm T}_{+1}\rangle \\\nonumber
|{\rm T}_z\rangle &=& |{\rm T}_0\rangle
\end{eqnarray}
$|{\rm S}\rangle$ and $|{\rm T}_m\rangle$ can also be written:
\begin{eqnarray} \label{eq:appdx5}
|{\rm S}\rangle      &=& \tfrac{1}{\sqrt{2}}\left[|\alpha_1\beta_2\rangle - |\beta_1\alpha_2\rangle\right] \\\nonumber
|{\rm T}_{+1}\rangle &=& |\alpha_1\alpha_2\rangle\\\nonumber
|{\rm T}_0\rangle    &=& \tfrac{1}{\sqrt{2}}\left[|\alpha_1\beta_2\rangle + |\beta_1\alpha_2\rangle\right] \\\nonumber
|{\rm T}_{-1}\rangle &=& |\beta_1\beta_2\rangle
\end{eqnarray}
where $|\alpha_j\rangle$ and $|\beta_j\rangle$ are defined by:
\begin{eqnarray} \label{eq:appdx6}
\langle\alpha_j|\hat{S}_{j,z}|\alpha_j\rangle &=& +\tfrac{1}{2} \\\nonumber
\langle\beta_j|\hat{S}_{j,z}|\beta_j\rangle   &=& -\tfrac{1}{2} \quad (j=1,2)
\end{eqnarray}
The axis system in which $\hat{S}_x$, $\hat{S}_y$ and $\hat{S}_z$ and $\hat{S}_{j,z}$ are defined may be chosen to be the `laboratory frame', in which the $z$-axis is commonly the direction of the applied magnetic field, or a `molecular frame', which could, for example, be the principal axis system of one of the hyperfine (HF) tensors. The triplet state $|{\rm T}_q\rangle$ $(q=x,y,z)$, Eq.~\eqref{eq:appdx3}, is spin-polarized in the $q=0$ principal plane within the molecule \cite{groo-mp-12-259}.

\paragraph{Initial state.}

The initial state of the radical pair spin system, $\hat{\rho}(0)$, is written as the direct product of the initial density operator for the two electron spins, $\hat{\rho}_0$, and identity operators for each of the nuclear spins ($i=1,2,\cdots$) to which the electrons are coupled:
\begin{eqnarray} \label{eq:appdx7}
\hat{\rho}(0) &=& \frac{1}{M}\hat{\rho}_0 \otimes \left\{\bigotimes_i\hat{\openone}_i\right\}
\end{eqnarray}
($M$ is the total dimension of the nuclear spin-space). It is assumed that the formation of the radical pair is not nuclear spin-dependent. In the absence of chemical reactivity and spin-decoherence, the probability that the radical pair is in a singlet state at time $t$ is given by the expectation value of the singlet projection operator, $\hat{P}^{\rm S}$:
\begin{eqnarray} \label{eq:appdx8}
\left\langle\hat{P}^{\rm S}\right\rangle\!(t) &=& \text{Tr}\left[\hat{\rho}(t)\hat{P}^{\rm S}\right] \\\nonumber
&=& \text{Tr}\left[{\rm e}^{-{\rm i}\hat{H}t}\hat{\rho}(0){\rm e}^{+{\rm i}\hat{H}t}\hat{P}^{\rm S}\right]
\end{eqnarray}
where $\hat{H}$ is the time-independent spin Hamiltonian and 
\begin{eqnarray} \label{eq:appdx9}
\hat{P}^{\rm S} &=& \left\{|{\rm S}\rangle\langle{\rm S}|\right\}\otimes\left\{\bigotimes_i\hat{\openone}_i\right\}
\end{eqnarray}

\paragraph{Minimal model.}

The principles of radical pair magnetoreception can be discussed using a simple model comprising two electron spins and one spin-\sfrac{1}{2} nucleus (e.g. $^1$H) with an axially anisotropic HF coupling to one of the electron spins:
\begin{eqnarray} \label{eq:appdx10}
\hat{H}_\text{hfi} &=& \sum_{q=x,y,z}A_{qq}\hat{S}_{1,q}\hat{I}_q \\\nonumber
 &=& a\sum_{q=x,y,z}\hat{S}_{1,q}\hat{I}_q + a\alpha\left(\hat{S}_{1,x}\hat{I}_x+\hat{S}_{1,y}\hat{I}_y-2\hat{S}_{1,z}\hat{I}_z\right)
\end{eqnarray}
$a$ is the isotropic part of the HF interaction (expressed as an angular frequency) and $\alpha$ is a dimensionless axiality parameter \cite{cint-cp-294-385}. Defined in this way, the HF interaction has cylindrical symmetry around the molecular $z$-axis.

Two cases are considered specifically: $\alpha=0$ (isotropic HF interaction) and $\alpha=-1$ (the HF interaction that results in the largest anisotropy in the reaction yield of this 3-spin system \cite{cai-pra-85-040304}).

The electron Zeeman interaction is included by means of the spin Hamiltonian:
\begin{eqnarray} \label{eq:appdx11}
\hat{H}_\text{Zeeman} &=& \omega\sum_{j=1,2}\left[\hat{S}_{j,z}\cos\theta+\hat{S}_{j,x}\sin\theta\right]
\end{eqnarray}
in which $\omega$ is the strength of the applied magnetic field (expressed as an angular frequency) and $\theta$ specifies its direction with respect to the symmetry axis ($z$) of the HF tensor. It is assumed that the $g$-tensors of the two radicals are identical and isotropic, and that the nuclear Zeeman interactions are negligible. Both are excellent approximations for organic radicals subject to the weak magnetic fields of interest here.

\begin{figure}
\includegraphics[width=3.375in]{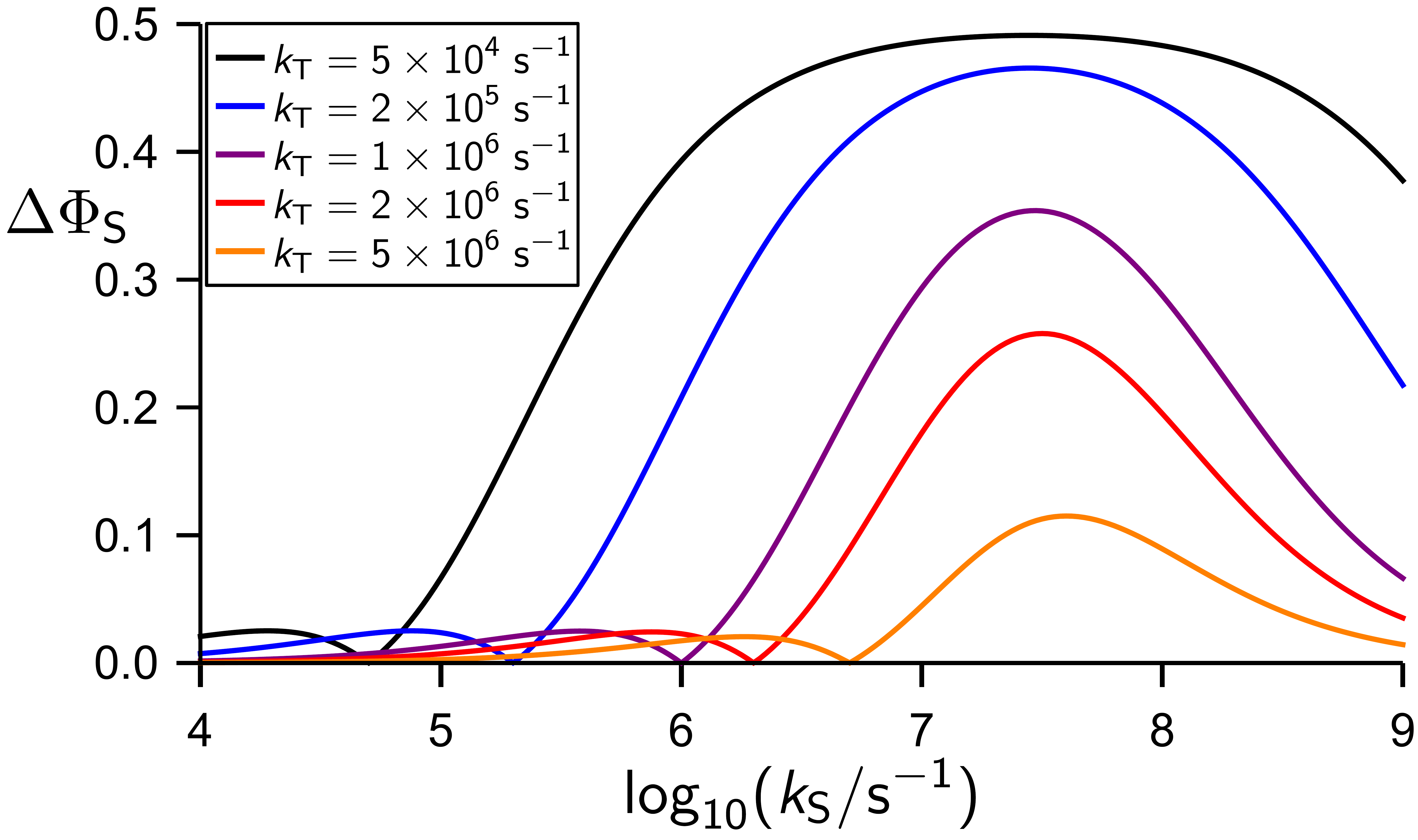}%
\caption{\label{fig:minimalrp} Reaction yield anisotropy, $\Delta\Phi_{\rm S}$, of the minimal radical pair model with a non-coherent initial state, $\hat{\rho}_0=\frac{1}{4}\hat{\openone}$. $\Delta\Phi_{\rm S}$ is shown as a function of the rate constants $k_{\rm S}$ and $k_{\rm T}$. $a = 1.0\;\rm mT$, $\alpha=-1$, $\omega = 50\;\rm \mu T$.}
\end{figure}

\begin{figure}
\includegraphics[width=3.375in]{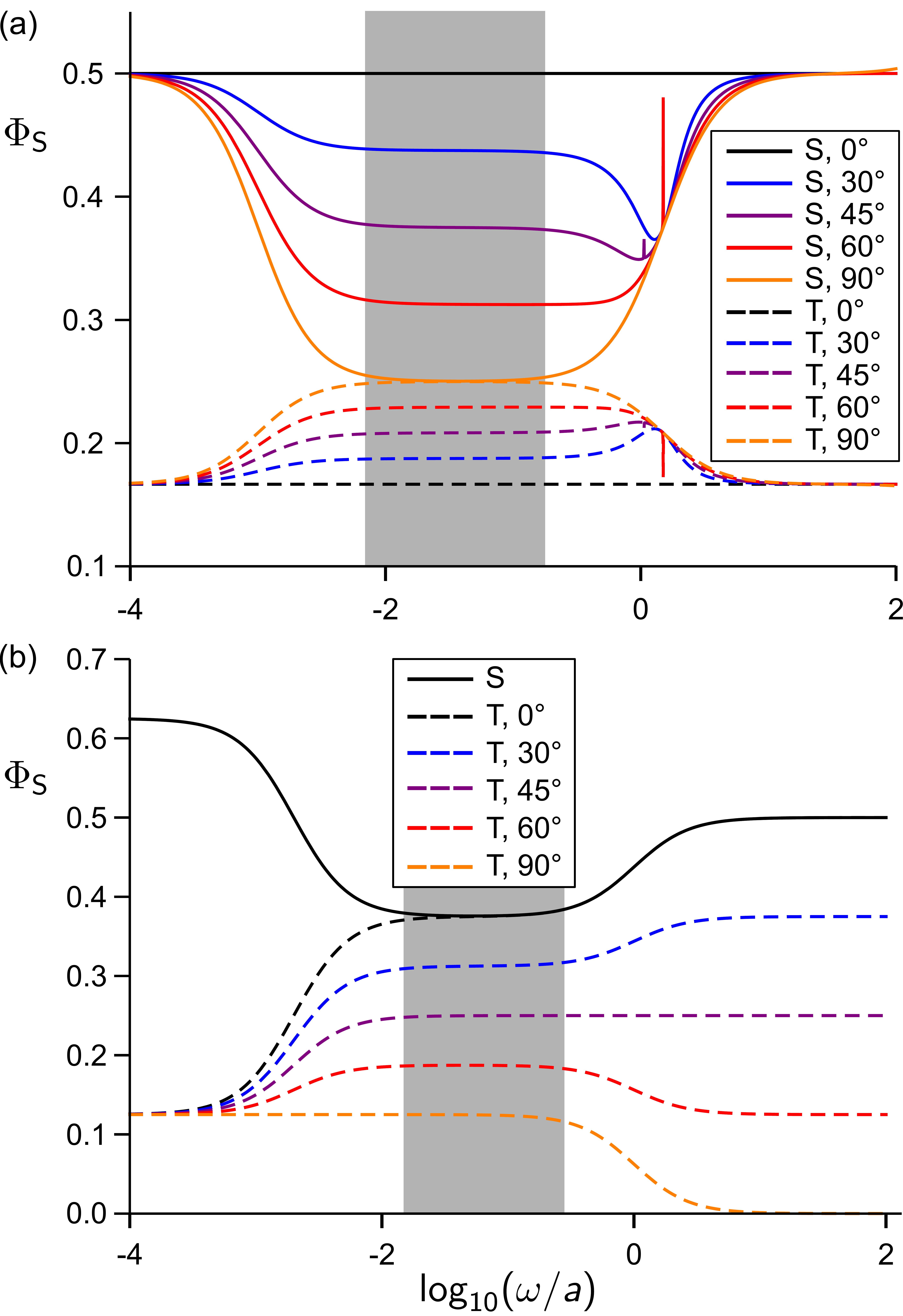}%
\caption{\label{fig:mfe} Magnetic field effect on the reaction yield of the minimal radical pair model. $k=10^{-3}a$. (a) Anisotropic HF interaction ($\alpha=-1$). S (solid lines) and T (dashed lines) denote the initial radical pair states   $\hat{\rho}_0=\hat{\rho}_0({\rm S})$ ($\mu=1$) and $\hat{\rho}_0=\hat{\rho}_0({\rm T})$ ($\mu=0$), respectively, in Eq. \eqref{eq:mixedstart}. The perturbation theory result in Eq. \eqref{eq:isotropic} is valid in the shaded region where $|a|\gg\omega\gg k$. The dependence of $\Phi_{\rm S}$ on the strength of the applied magnetic field ($\omega/a$) is shown for various angles between the symmetry axis of the HF tensor and the magnetic field vector. The sharp features near $\log_{10}(\omega/a) = 0.3$ arise from level anti-crossings \cite{timm-cpl-334-387}.
(b) Isotropic HF interaction ($\alpha = 0$). S (solid line) and T (dashed lines) denote $\hat{\rho}_0=|{\rm S}\rangle\langle{\rm S}|$ ($\eta=1$) and $\hat{\rho}_0=|{\rm T}_z\rangle\langle{\rm T}_z|$ ($\eta=0$), respectively, in Eq. \eqref{eq:lincomb}. The perturbation theory result in Eq. \eqref{eq:anisotropic} is valid in the shaded region where $|a|\gg\omega\gg k$. The dependence of $\Phi_{\rm S}$ on the strength of the applied magnetic field ($\omega / a$) is shown for various angles between the triplet alignment axis and the magnetic field vector.}
\end{figure}

To account for the chemical reactivity of the radical pair within the minimal model, we use the `exponential model' \cite{timm-mp-95-71} in which singlet and triplet states react spin-selectively with the same first-order rate constant, $k_{\rm S}=k_{\rm T}=k$, to form distinct products. Although unlikely to be strictly valid for any real magnetoreceptor, this approximation simplifies the algebra without distorting the underlying physics. The yield of the chemical product formed via the \emph{singlet} pathway, and its anisotropy, are calculated as \cite{timm-mp-95-71}:
\begin{eqnarray} \label{eq:appdx12}
\Phi_{\rm S} &=&k\int_0^\infty\left\langle\hat{P}^{\rm S}\right\rangle\!(t){\rm e}^{-kt}{\rm d}t
\\ \label{eq:appdx13}
\Delta\Phi_{\rm S} &=& \text{max}\left(\Phi_{\rm S}\right) - \text{min}\left(\Phi_{\rm S}\right)
\end{eqnarray}
so that $0\le\Phi_{\rm S}\le 1$. The corresponding yield for the triplet reaction channel is simply $1-\Phi_{\rm S}$. $\Phi_{\rm S}$ is referred to as the reaction yield and $\Delta\Phi_{\rm S}$ as the reaction yield anisotropy. The variation of $\Phi_{\rm S}$ with the orientation of the radical pair with respect to an external magnetic field forms the basis of the compass mechanism \cite{schu-zpc-111-1}. 

When $k_{\rm S}\ne k_{\rm T}$, the calculation of $\Phi_{\rm S}$ is performed in Liouville space:
\sffamily
\begin{eqnarray} \label{eq:appdx14}
\Phi_{\rm S} &=& k_{\rm S}\left\langle\hat{P}^{\rm S}|\hat{\hat{L}}^{-1}|\hat{\rho}(0)\right\rangle
\\ \label{eq:appdx15}
\hat{\hat{L}} &=& {\rm i}\left(\hat{H}\otimes\hat{\openone}_8-\hat{\openone}_8\otimes\hat{H}^{\sf T}\right) \\\nonumber
&& + \tfrac{1}{2}k_{\rm S}\left(\hat{P}^{\rm S}\otimes\hat{\openone}_8 + \hat{\openone}_8\otimes\hat{P}^{\rm S}\right) \\\nonumber
&& + \tfrac{1}{2}k_{\rm T}\left(\hat{P}^{\rm T}\otimes\hat{\openone}_8 + \hat{\openone}_8\otimes\hat{P}^{\rm T}\right)
\end{eqnarray}
\normalfont
where $\hat{\openone}_8$ is the identity operator in the 8-dimensional spin-space and $\hat{P}^{\rm T}=\hat{\openone}_8-\hat{P}^{\rm S}$.

Fig.~\ref{fig:minimalrp} shows simulations for the minimal radical pair with an anisotropic HF coupling and an initial state:
\begin{eqnarray} \label{eq:appdx16}
\hat{\rho}_0 &=& \tfrac{1}{4}\hat{\rho}_0({\rm S})+\tfrac{3}{4}\hat{\rho}_0({\rm T}) = \tfrac{1}{4}\hat{\openone}
\end{eqnarray}
$\Delta\Phi_{\rm S}$ is non-zero except when $k_{\rm S}=k_{\rm T}$. The radical pair can exhibit magnetic compass properties even when its initial electron spin state is neither entangled nor coherent. The coherence arises during the spin evolution as a result of the differential reactivity of the singlet and triplet states.

\paragraph{Perturbation theory.}

To obtain estimates of the maximum possible magnetic responses within the minimal model, we use a perturbative approach \cite{timm-cpl-334-387,cint-cp-294-385}, appropriate for weak applied magnetic fields and long-lived radical pairs. This approximation is valid when $|a|\gg\omega\gg k$. These conditions are not unrealistic: HF interactions are of the order of $10^8\;\rm rad\; s^{-1}$ ($\approx 500\;\rm\mu T$), the geomagnetic field is roughly $10^7\;\rm rad\; s^{-1}$ ($\approx 50\;\rm\mu T$), and plausible values of the rate constant $k$ are $10^5-10^6\;\rm s^{-1}$ \cite{rodg-pnas-106-353}. 

Eqs~\eqref{eq:isotropic} and \eqref{eq:anisotropic} were verified by the exact numerical simulations, the results of which are shown in Fig. \ref{fig:mfe}.

\paragraph{Multinuclear radical pairs.}

\begin{figure}
\includegraphics[width=3in]{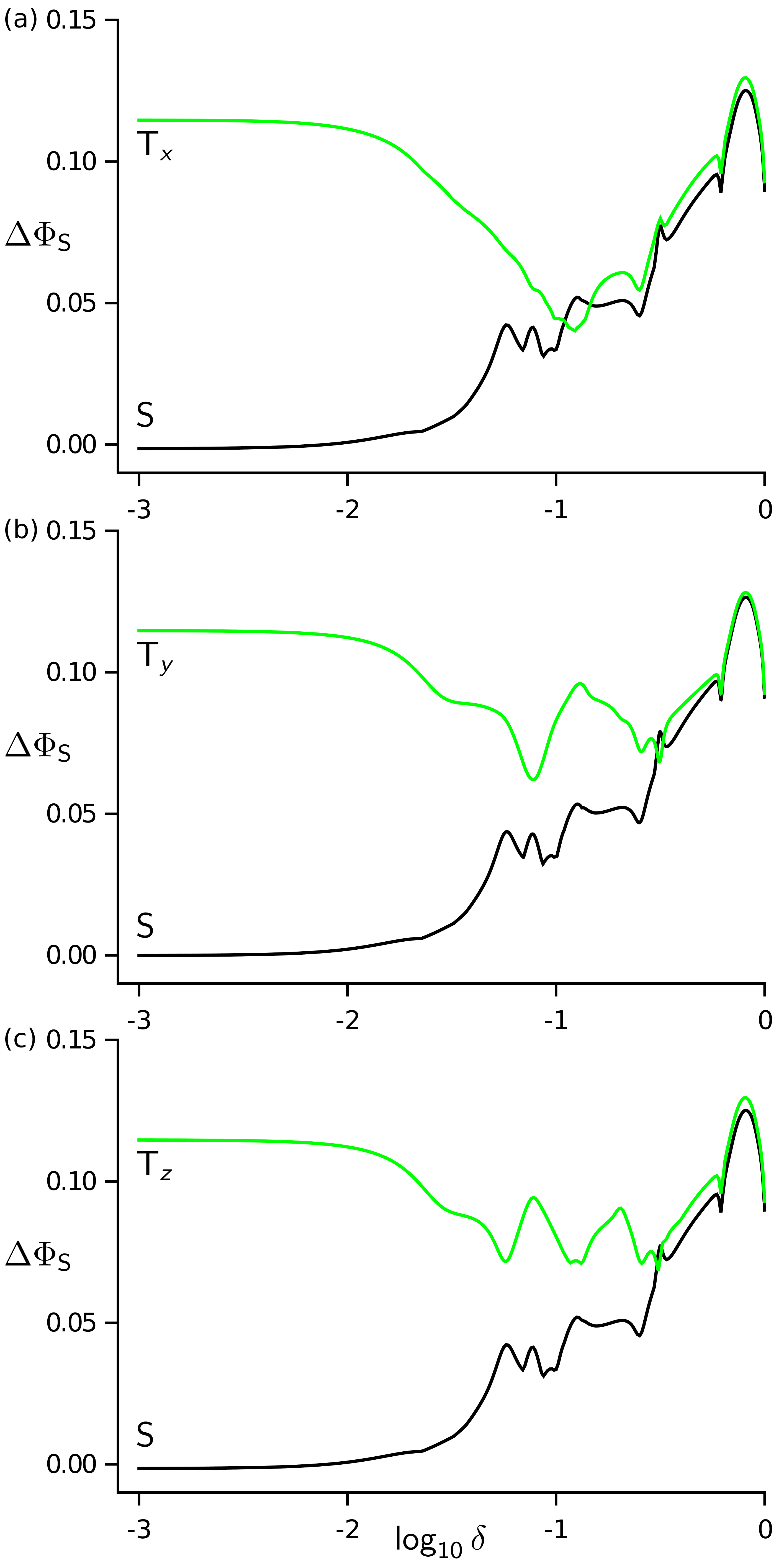}%
\caption{\label{fig:anisoappdx} Reaction yield anisotropy, $\Delta\Phi_{\rm S}$, calculated for a radical pair in which one radical contains a $^1$H nucleus (spin-\sfrac{1}{2}) and a $^{14}$N nucleus (spin-1). $k=10^6\;\rm s^{-1}$ and $\omega=50\;\rm\mu T$. The HF coupling parameters (in mT) are as in the caption for Fig. \ref{fig:rp11}. $\hat{\rho}_0=|{\rm S}\rangle\langle{\rm S}|$ (black). $\hat{\rho}_0=|{\rm T}_x\rangle\langle{\rm T}_x|$ (a, green), $\hat{\rho}_0=|{\rm T}_y\rangle\langle{\rm T}_y|$ (b, green), $\hat{\rho}_0=|{\rm T}_z\rangle\langle{\rm T}_z|$ (c, green).}
\end{figure}

In the general case, the HF component of the spin Hamiltonian has the form:
\begin{eqnarray} \label{eq:appdx17}
\hat{H}_\text{hfi} &=& \sum_{j=1,2}\sum_k\left[\hat{\mathbf{S}}_j\cdot\mathbf{A}_{jk}\cdot\hat{\mathbf{I}}_k\right] \\\nonumber
 &=& \sum_{j=1,2}\sum_k\left[a_{jk}\hat{\mathbf{S}}_j\cdot\hat{\mathbf{I}}_k + \hat{\mathbf{S}}_j\cdot\mathbf{T}_{jk}\cdot\hat{\mathbf{I}}_k\right]
\end{eqnarray}
where $a_{jk}$, $\mathbf{T}_{jk}$ and $\mathbf{A}_{jk}$ are, respectively, the isotropic HF coupling constant, the anisotropic HF tensor and the total HF tensor for nucleus $k$ coupled to the electron in radical $j$.

The Zeeman term is:
\begin{eqnarray} \label{eq:appdx18}
\hat{H}_\text{Zeeman} &=& \omega\sum_{j=1,2}\left[\hat{S}_{j,x}\sin\theta\cos\phi\right. \\\nonumber 
&&\qquad\qquad + \left.\hat{S}_{j,y}\sin\theta\sin\phi + \hat{S}_{j,z}\cos\theta\right]
\end{eqnarray}
where $\theta$ and $\phi$ specify the direction of the field in the molecular axis system. 

Fig.~\ref{fig:anisoappdx} shows, for completeness, versions of Fig.~\ref{fig:rp11} in which the three initial triplet states are compared with $|{\rm S}\rangle\langle{\rm S}|$.

\paragraph{Concurrence.}

To quantify the entanglement of the various initial electron spin states $\hat{\rho}_0$, we use the `concurrence' proposed by Wootters for a two-qubit density operator \cite{woot-prl-80-2245}:
\begin{eqnarray} \label{eq:appdx19}
C(\hat{\rho}_0) &=& \text{max}\left\{0,\lambda_1-\lambda_2-\lambda_3-\lambda_4\right\}
\end{eqnarray}
where the $\lambda_i$ are the non-negative real square roots of the eigenvalues, in decreasing order, of:
\begin{eqnarray} \label{eq:appdx20}
\hat{\rho}_0\left(\hat{\sigma}_y\otimes\hat{\sigma}_y\right)\hat{\rho}_0^\ast\left(\hat{\sigma}_y\otimes\hat{\sigma}_y\right)
\end{eqnarray}
in which $\hat{\sigma}_y$ is twice the $\hat{S}_y$ operator for a single electron spin, and $\hat{\rho}_0^\ast$ is the complex conjugate of $\hat{\rho}_0$.

\paragraph{General initial conditions.}

Within the minimal model, the most general initial state consistent with both Eqs~\eqref{eq:mixedstart} and \eqref{eq:molframe} is:
\begin{eqnarray} \label{eq:appdx21}
\hat{\rho}_0 &=& \varepsilon|{\rm S}\rangle\langle{\rm S}| + (1-\varepsilon)\sum_{q=x,y,z}p_q|{\rm T}_q\rangle\langle{\rm T}_q|
\end{eqnarray}
with $0\le\varepsilon\le 1$, $p_x,p_y,p_z \ge 0$ and $p_x+p_y+p_z=1$. Some such states are entangled and some are not. Almost all are anisotropic and may lead to compass behaviour even when the spin-Hamiltonian is isotropic. For the minimal model, with an isotropic HF coupling and $|a|\gg\omega\gg k$, the initial state in Eq.~\eqref{eq:appdx21} leads to:
\begin{eqnarray} \label{eq:appdx22}
\Phi_{\rm S} &=& \tfrac{1}{8}(2\varepsilon+1)+\tfrac{1}{4}(1-\varepsilon)\\\nonumber
&& \left(p_x\sin^2\theta\cos^2\phi + p_y\sin^2\theta\sin^2\phi + p_z\cos^2\theta\right)
\\
\Delta\Phi_{\rm S} &=& \tfrac{1}{4}(1-\varepsilon)(p_\text{max}-p_\text{min})
\end{eqnarray}
where $p_\text{max} = \text{max}\{p_x,p_y,p_z\}$ and $p_\text{min} = \text{min}\{p_x,p_y,p_z\}$. The maximum anisotropy is obtained when $\varepsilon = 0$, $p_\text{max}=1$ and $p_\text{min}=0$ (giving $\Delta\Phi_{\rm S}=\frac{1}{4}$). 

The concurrence $C(\hat{\rho}_0)$ of the state in Eq. \eqref{eq:appdx21} is:
\begin{eqnarray} \label{eq:appdx23}
2\varepsilon-1 \quad &&\text{when}\quad \varepsilon > \tfrac{1}{2}
\\
2p_\text{max}(1-\varepsilon)-1 \quad &&\text{when}\quad \varepsilon \le \tfrac{1}{2} \\\nonumber
&&\quad\text{and}\quad p_\text{max}\ge\tfrac{1}{2(1-\varepsilon)}
\\
0 \quad && \text{otherwise}
\end{eqnarray}

\paragraph{Laboratory-frame polarization}

Although not relevant for a geomagnetic compass sensor, radical pairs can be created with large laboratory-frame polarizations, i.e. with unequal populations of the triplet eigenstates in a strong magnetic field ($|{\rm T}_m\rangle$ ($m=0,\pm 1$), as defined by Eq.~\eqref{eq:appdx2}, with the $z$-axis being the direction of a strong applied magnetic field). For example, the Triplet Mechanism of Chemically Induced Dynamic Electron Polarization can result in large polarizations for radical pairs produced by triplet states formed by anisotropic intersystem crossing \cite{atki-mp-27-1633,hore-cpl-69-563}. Another example, which also requires the electron spins to be quantized by strong electron Zeeman interactions, is seen in the EPR spectra of spin-correlated radical pairs with non-zero electron-electron exchange and/or dipolar interactions \cite{buck-cpl-135-307,hore-cpl-137-495}.

\begin{table}
\caption{\label{tab:fadhf} HF data for N5 and H5 in FADH$^\bullet$.}
\begin{ruledtabular}
\begin{tabular}{crrrrr}
Nucleus & $a$ / mT & $T_{jj}$ / mT & \multicolumn{3}{c}{Principal axes}
\\
N5 &  $0.393$ & $-0.498$ &  $0.4380$ &  $0.8655$ & $-0.2432$
\\
   &          & $-0.492$ &  $0.8981$ & $-0.4097$ & $0.1595$
\\
   &          &  $0.990$ & $-0.0384$ &  $0.2883$ & $0.9568$
\\
H5 & $-0.769$ & $-0.616$ &  $0.9819$ &  $0.1883$ & $-0.0203$
\\
   &          & $-0.168$ & $-0.0348$ &  $0.2850$ &  $0.9579$
\\
   &          &  $0.784$ & $-0.1861$ &  $0.9398$ & $-0.2864$
\end{tabular}
\end{ruledtabular}

\vspace*{1em}

\includegraphics[width=1.5in]{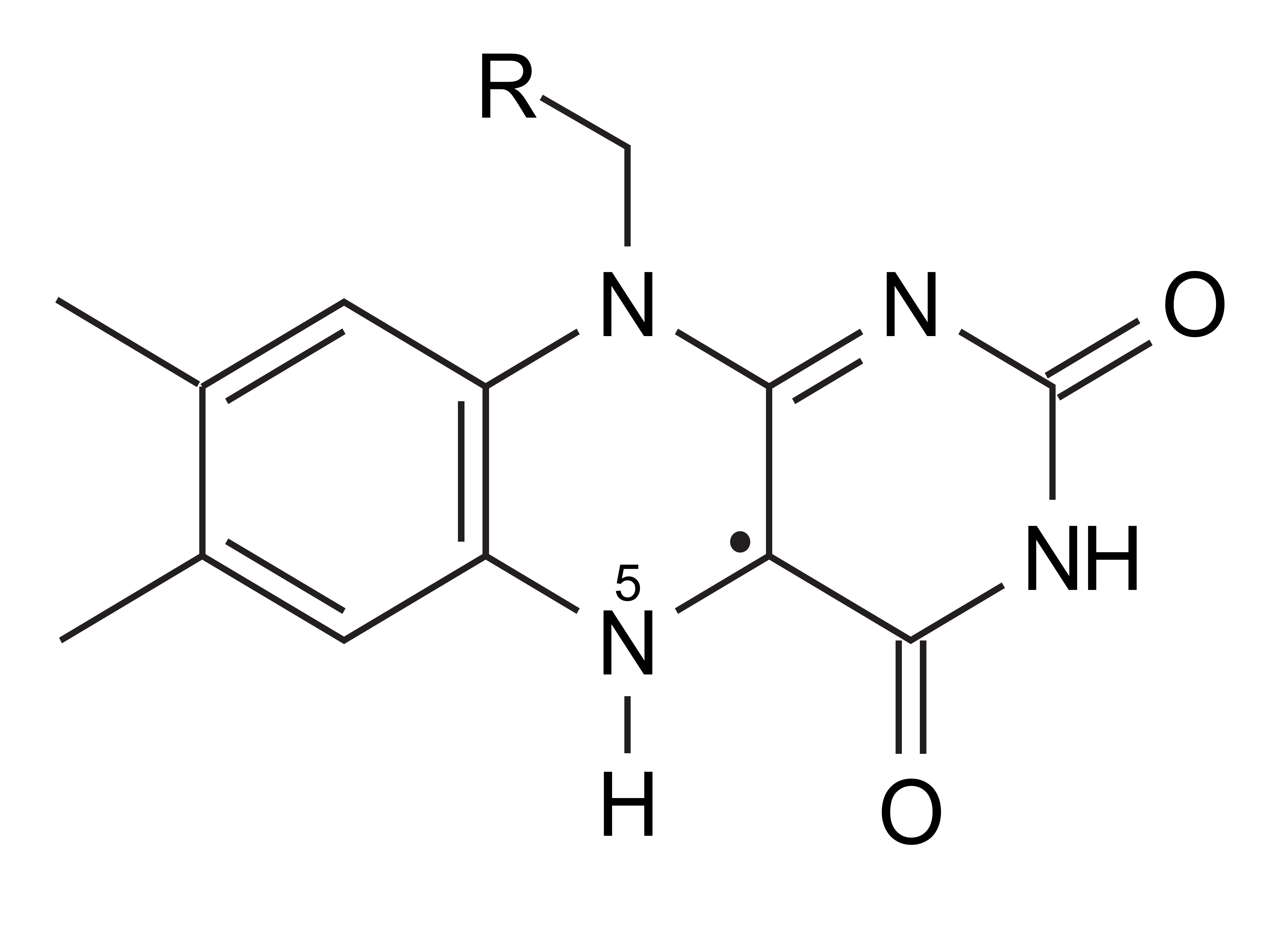}
\end{table}

\paragraph{Intersystem crossing.}

The implications of anisotropic intersystem crossing for the behaviour and properties of radical pairs is discussed in detail by Steiner and Ulrich \cite{stei-cr-89-51} (pp. 109--112). In most cases, as here, intersystem crossing is assumed to be independent of HF-coupled nuclear spins. However, Kothe \emph{et al.} \cite{koth-jpcb-114-14755}, in an elegant study of quantum oscillations in an organic triplet state, have shown that the nuclei are in fact involved and that the appropriate molecular-frame triplet basis states are eigenstates of the combined zero-field and hyperfine Hamiltonians. We do not consider this possibility here.

\paragraph{FADH$^\bullet$ radical.}

The HF interactions for FADH$^\bullet$ used to calculate the reaction yield anisotropies shown in Fig.~\ref{fig:rp40} are based on the data in Table~\ref{tab:fadhf} \cite{webe-jacs-123-3790}.


\bibliography{abbrjrnl,entanglement}

\providecommand{\noopsort}[1]{}\providecommand{\singleletter}[1]{#1}%
\begin{thebibliography}{35}%
\makeatletter
\providecommand \@ifxundefined [1]{%
 \@ifx{#1\undefined}
}%
\providecommand \@ifnum [1]{%
 \ifnum #1\expandafter \@firstoftwo
 \else \expandafter \@secondoftwo
 \fi
}%
\providecommand \@ifx [1]{%
 \ifx #1\expandafter \@firstoftwo
 \else \expandafter \@secondoftwo
 \fi
}%
\providecommand \natexlab [1]{#1}%
\providecommand \enquote  [1]{``#1''}%
\providecommand \bibnamefont  [1]{#1}%
\providecommand \bibfnamefont [1]{#1}%
\providecommand \citenamefont [1]{#1}%
\providecommand \href@noop [0]{\@secondoftwo}%
\providecommand \href [0]{\begingroup \@sanitize@url \@href}%
\providecommand \@href[1]{\@@startlink{#1}\@@href}%
\providecommand \@@href[1]{\endgroup#1\@@endlink}%
\providecommand \@sanitize@url [0]{\catcode `\\12\catcode `\$12\catcode
  `\&12\catcode `\#12\catcode `\^12\catcode `\_12\catcode `\%12\relax}%
\providecommand \@@startlink[1]{}%
\providecommand \@@endlink[0]{}%
\providecommand \url  [0]{\begingroup\@sanitize@url \@url }%
\providecommand \@url [1]{\endgroup\@href {#1}{\urlprefix }}%
\providecommand \urlprefix  [0]{URL }%
\providecommand \Eprint [0]{\href }%
\providecommand \doibase [0]{http://dx.doi.org/}%
\providecommand \selectlanguage [0]{\@gobble}%
\providecommand \bibinfo  [0]{\@secondoftwo}%
\providecommand \bibfield  [0]{\@secondoftwo}%
\providecommand \translation [1]{[#1]}%
\providecommand \BibitemOpen [0]{}%
\providecommand \bibitemStop [0]{}%
\providecommand \bibitemNoStop [0]{.\EOS\space}%
\providecommand \EOS [0]{\spacefactor3000\relax}%
\providecommand \BibitemShut  [1]{\csname bibitem#1\endcsname}%
\let\auto@bib@innerbib\@empty
\bibitem [{\citenamefont {Winklhofer}\ and\ \citenamefont
  {Kirschvink}(2010)}]{wink-jrsif-7-S273}%
  \BibitemOpen
  \bibfield  {author} {\bibinfo {author} {\bibfnamefont {M.}~\bibnamefont
  {Winklhofer}}\ and\ \bibinfo {author} {\bibfnamefont {J.~L.}\ \bibnamefont
  {Kirschvink}},\ }\href@noop {} {\bibfield  {journal} {\bibinfo  {journal}
  {J.\ R.\ Soc.\ Interface}\ }\textbf {\bibinfo {volume} {7}},\ \bibinfo
  {pages} {S273} (\bibinfo {year} {2010})}\BibitemShut {NoStop}%
\bibitem [{\citenamefont {Schulten}\ \emph {et~al.}(1978)\citenamefont
  {Schulten}, \citenamefont {Swenberg},\ and\ \citenamefont
  {Weller}}]{schu-zpc-111-1}%
  \BibitemOpen
  \bibfield  {author} {\bibinfo {author} {\bibfnamefont {K.}~\bibnamefont
  {Schulten}}, \bibinfo {author} {\bibfnamefont {C.~E.}\ \bibnamefont
  {Swenberg}}, \ and\ \bibinfo {author} {\bibfnamefont {A.}~\bibnamefont
  {Weller}},\ }\href@noop {} {\bibfield  {journal} {\bibinfo  {journal} {Z.\
  Phys.\ Chemie}\ }\textbf {\bibinfo {volume} {111}},\ \bibinfo {pages} {1}
  (\bibinfo {year} {1978})}\BibitemShut {NoStop}%
\bibitem [{\citenamefont {Ritz}\ \emph {et~al.}(2000)\citenamefont {Ritz},
  \citenamefont {Adem},\ and\ \citenamefont {Schulten}}]{ritz-bj-78-707}%
  \BibitemOpen
  \bibfield  {author} {\bibinfo {author} {\bibfnamefont {T.}~\bibnamefont
  {Ritz}}, \bibinfo {author} {\bibfnamefont {S.}~\bibnamefont {Adem}}, \ and\
  \bibinfo {author} {\bibfnamefont {K.}~\bibnamefont {Schulten}},\ }\href@noop
  {} {\bibfield  {journal} {\bibinfo  {journal} {Biophys.\ J.}\ }\textbf
  {\bibinfo {volume} {78}},\ \bibinfo {pages} {707} (\bibinfo {year}
  {2000})}\BibitemShut {NoStop}%
\bibitem [{\citenamefont {Maeda}\ \emph {et~al.}(2012)\citenamefont {Maeda},
  \citenamefont {Robinson}, \citenamefont {Henbest}, \citenamefont {Hogben},
  \citenamefont {Biskup}, \citenamefont {Ahmad}, \citenamefont {Schleicher},
  \citenamefont {Weber}, \citenamefont {Timmel},\ and\ \citenamefont
  {Hore}}]{maed-pnas-109-4774}%
  \BibitemOpen
  \bibfield  {author} {\bibinfo {author} {\bibfnamefont {K.}~\bibnamefont
  {Maeda}}, \bibinfo {author} {\bibfnamefont {A.~J.}\ \bibnamefont {Robinson}},
  \bibinfo {author} {\bibfnamefont {K.~B.}\ \bibnamefont {Henbest}}, \bibinfo
  {author} {\bibfnamefont {H.~J.}\ \bibnamefont {Hogben}}, \bibinfo {author}
  {\bibfnamefont {T.}~\bibnamefont {Biskup}}, \bibinfo {author} {\bibfnamefont
  {M.}~\bibnamefont {Ahmad}}, \bibinfo {author} {\bibfnamefont
  {E.}~\bibnamefont {Schleicher}}, \bibinfo {author} {\bibfnamefont
  {S.}~\bibnamefont {Weber}}, \bibinfo {author} {\bibfnamefont {C.~R.}\
  \bibnamefont {Timmel}}, \ and\ \bibinfo {author} {\bibfnamefont {P.~J.}\
  \bibnamefont {Hore}},\ }\href@noop {} {\bibfield  {journal} {\bibinfo
  {journal} {Proc.\ Natl.\ Acad.\ Sci.\ USA}\ }\textbf {\bibinfo {volume}
  {109}},\ \bibinfo {pages} {4774} (\bibinfo {year} {2012})}\BibitemShut
  {NoStop}%
\bibitem [{\citenamefont {Rodgers}\ and\ \citenamefont
  {Hore}(2009)}]{rodg-pnas-106-353}%
  \BibitemOpen
  \bibfield  {author} {\bibinfo {author} {\bibfnamefont {C.~T.}\ \bibnamefont
  {Rodgers}}\ and\ \bibinfo {author} {\bibfnamefont {P.~J.}\ \bibnamefont
  {Hore}},\ }\href@noop {} {\bibfield  {journal} {\bibinfo  {journal} {Proc.\
  Natl.\ Acad.\ Sci.\ USA}\ }\textbf {\bibinfo {volume} {106}},\ \bibinfo
  {pages} {353} (\bibinfo {year} {2009})}\BibitemShut {NoStop}%
\bibitem [{\citenamefont {Liedvogel}\ and\ \citenamefont
  {Mouritsen}(2010)}]{lied-jrsif-7-S147}%
  \BibitemOpen
  \bibfield  {author} {\bibinfo {author} {\bibfnamefont {M.}~\bibnamefont
  {Liedvogel}}\ and\ \bibinfo {author} {\bibfnamefont {H.}~\bibnamefont
  {Mouritsen}},\ }\href@noop {} {\bibfield  {journal} {\bibinfo  {journal} {J.\
  R.\ Soc.\ Interface}\ }\textbf {\bibinfo {volume} {7}},\ \bibinfo {pages}
  {S147} (\bibinfo {year} {2010})}\BibitemShut {NoStop}%
\bibitem [{\citenamefont {Ritz}(2011)}]{ritz-pchem-3-262}%
  \BibitemOpen
  \bibfield  {author} {\bibinfo {author} {\bibfnamefont {T.}~\bibnamefont
  {Ritz}},\ }\href@noop {} {\bibfield  {journal} {\bibinfo  {journal} {Procedia
  Chem.}\ }\textbf {\bibinfo {volume} {3}},\ \bibinfo {pages} {262} (\bibinfo
  {year} {2011})}\BibitemShut {NoStop}%
\bibitem [{\citenamefont {Mouritsen}\ and\ \citenamefont
  {Hore}(2012)}]{mour-conb-22-343}%
  \BibitemOpen
  \bibfield  {author} {\bibinfo {author} {\bibfnamefont {H.}~\bibnamefont
  {Mouritsen}}\ and\ \bibinfo {author} {\bibfnamefont {P.~J.}\ \bibnamefont
  {Hore}},\ }\href@noop {} {\bibfield  {journal} {\bibinfo  {journal} {Curr.\
  Opin.\ Neurobiol.}\ }\textbf {\bibinfo {volume} {22}},\ \bibinfo {pages}
  {343} (\bibinfo {year} {2012})}\BibitemShut {NoStop}%
\bibitem [{\citenamefont {Ball}(2011)}]{ball-n-474-242}%
  \BibitemOpen
  \bibfield  {author} {\bibinfo {author} {\bibfnamefont {P.}~\bibnamefont
  {Ball}},\ }\href@noop {} {\bibfield  {journal} {\bibinfo  {journal} {Nature}\
  }\textbf {\bibinfo {volume} {474}},\ \bibinfo {pages} {272} (\bibinfo {year}
  {2011})}\BibitemShut {NoStop}%
\bibitem [{\citenamefont {Cai}\ \emph {et~al.}(2010)\citenamefont {Cai},
  \citenamefont {Guerreschi},\ and\ \citenamefont
  {Briegel}}]{cai-prl-104-220502}%
  \BibitemOpen
  \bibfield  {author} {\bibinfo {author} {\bibfnamefont {J.}~\bibnamefont
  {Cai}}, \bibinfo {author} {\bibfnamefont {G.~G.}\ \bibnamefont {Guerreschi}},
  \ and\ \bibinfo {author} {\bibfnamefont {H.~J.}\ \bibnamefont {Briegel}},\
  }\href@noop {} {\bibfield  {journal} {\bibinfo  {journal} {Phys.\ Rev.\
  Lett.}\ }\textbf {\bibinfo {volume} {104}},\ \bibinfo {pages} {220502}
  (\bibinfo {year} {2010})}\BibitemShut {NoStop}%
\bibitem [{\citenamefont {Gauger}\ \emph {et~al.}(2011)\citenamefont {Gauger},
  \citenamefont {Rieper}, \citenamefont {Morton}, \citenamefont {Benjamin},\
  and\ \citenamefont {Vedral}}]{gaug-prl-106-040503}%
  \BibitemOpen
  \bibfield  {author} {\bibinfo {author} {\bibfnamefont {E.~M.}\ \bibnamefont
  {Gauger}}, \bibinfo {author} {\bibfnamefont {E.}~\bibnamefont {Rieper}},
  \bibinfo {author} {\bibfnamefont {J.~J.~L.}\ \bibnamefont {Morton}}, \bibinfo
  {author} {\bibfnamefont {S.~C.}\ \bibnamefont {Benjamin}}, \ and\ \bibinfo
  {author} {\bibfnamefont {V.}~\bibnamefont {Vedral}},\ }\href@noop {}
  {\bibfield  {journal} {\bibinfo  {journal} {Phys.\ Rev.\ Lett.}\ }\textbf
  {\bibinfo {volume} {106}},\ \bibinfo {pages} {040503} (\bibinfo {year}
  {2011})}\BibitemShut {NoStop}%
\bibitem [{\citenamefont {Cai}\ \emph {et~al.}(2012{\natexlab{a}})\citenamefont
  {Cai}, \citenamefont {Ai}, \citenamefont {Quan},\ and\ \citenamefont
  {Sun}}]{cai-pra-85-022315}%
  \BibitemOpen
  \bibfield  {author} {\bibinfo {author} {\bibfnamefont {C.~Y.}\ \bibnamefont
  {Cai}}, \bibinfo {author} {\bibfnamefont {Q.}~\bibnamefont {Ai}}, \bibinfo
  {author} {\bibfnamefont {H.~T.}\ \bibnamefont {Quan}}, \ and\ \bibinfo
  {author} {\bibfnamefont {C.~P.}\ \bibnamefont {Sun}},\ }\href@noop {}
  {\bibfield  {journal} {\bibinfo  {journal} {Phys.\ Rev.\ A}\ }\textbf
  {\bibinfo {volume} {85}},\ \bibinfo {pages} {022315} (\bibinfo {year}
  {2012}{\natexlab{a}})}\BibitemShut {NoStop}%
\bibitem [{\citenamefont {Cai}\ \emph {et~al.}(2012{\natexlab{b}})\citenamefont
  {Cai}, \citenamefont {Caruso},\ and\ \citenamefont
  {Plenio}}]{cai-pra-85-040304}%
  \BibitemOpen
  \bibfield  {author} {\bibinfo {author} {\bibfnamefont {J.}~\bibnamefont
  {Cai}}, \bibinfo {author} {\bibfnamefont {F.}~\bibnamefont {Caruso}}, \ and\
  \bibinfo {author} {\bibfnamefont {M.~B.}\ \bibnamefont {Plenio}},\
  }\href@noop {} {\bibfield  {journal} {\bibinfo  {journal} {Phys.\ Rev.\ A}\
  }\textbf {\bibinfo {volume} {85}},\ \bibinfo {pages} {040304} (\bibinfo
  {year} {2012}{\natexlab{b}})}\BibitemShut {NoStop}%
\bibitem [{\citenamefont {Cintolesi}\ \emph {et~al.}(2003)\citenamefont
  {Cintolesi}, \citenamefont {Ritz}, \citenamefont {Kay}, \citenamefont
  {Timmel},\ and\ \citenamefont {Hore}}]{cint-cp-294-385}%
  \BibitemOpen
  \bibfield  {author} {\bibinfo {author} {\bibfnamefont {F.}~\bibnamefont
  {Cintolesi}}, \bibinfo {author} {\bibfnamefont {T.}~\bibnamefont {Ritz}},
  \bibinfo {author} {\bibfnamefont {C.~W.~M.}\ \bibnamefont {Kay}}, \bibinfo
  {author} {\bibfnamefont {C.~R.}\ \bibnamefont {Timmel}}, \ and\ \bibinfo
  {author} {\bibfnamefont {P.~J.}\ \bibnamefont {Hore}},\ }\href@noop {}
  {\bibfield  {journal} {\bibinfo  {journal} {Chem.\ Phys.}\ }\textbf {\bibinfo
  {volume} {294}},\ \bibinfo {pages} {385} (\bibinfo {year}
  {2003})}\BibitemShut {NoStop}%
\bibitem [{\citenamefont {Maeda}\ \emph {et~al.}(2008)\citenamefont {Maeda},
  \citenamefont {Henbest}, \citenamefont {Cintolesi}, \citenamefont {Kuprov},
  \citenamefont {Rodgers}, \citenamefont {Liddell}, \citenamefont {Gust},
  \citenamefont {Timmel},\ and\ \citenamefont {Hore}}]{maed-n-453-387}%
  \BibitemOpen
  \bibfield  {author} {\bibinfo {author} {\bibfnamefont {K.}~\bibnamefont
  {Maeda}}, \bibinfo {author} {\bibfnamefont {K.~B.}\ \bibnamefont {Henbest}},
  \bibinfo {author} {\bibfnamefont {F.}~\bibnamefont {Cintolesi}}, \bibinfo
  {author} {\bibfnamefont {I.}~\bibnamefont {Kuprov}}, \bibinfo {author}
  {\bibfnamefont {C.~T.}\ \bibnamefont {Rodgers}}, \bibinfo {author}
  {\bibfnamefont {P.~A.}\ \bibnamefont {Liddell}}, \bibinfo {author}
  {\bibfnamefont {D.}~\bibnamefont {Gust}}, \bibinfo {author} {\bibfnamefont
  {C.~R.}\ \bibnamefont {Timmel}}, \ and\ \bibinfo {author} {\bibfnamefont
  {P.~J.}\ \bibnamefont {Hore}},\ }\href@noop {} {\bibfield  {journal}
  {\bibinfo  {journal} {Nature}\ }\textbf {\bibinfo {volume} {453}},\ \bibinfo
  {pages} {387} (\bibinfo {year} {2008})}\BibitemShut {NoStop}%
\bibitem [{\citenamefont {Weber}\ \emph {et~al.}(2010)\citenamefont {Weber},
  \citenamefont {Biskup}, \citenamefont {Okafuji}, \citenamefont {Marino},
  \citenamefont {Berthold}, \citenamefont {Link}, \citenamefont {Hitomi},
  \citenamefont {Getzoff}, \citenamefont {Schleicher},\ and\ \citenamefont
  {Norris}}]{webe-jpcb-114-14745}%
  \BibitemOpen
  \bibfield  {author} {\bibinfo {author} {\bibfnamefont {S.}~\bibnamefont
  {Weber}}, \bibinfo {author} {\bibfnamefont {T.}~\bibnamefont {Biskup}},
  \bibinfo {author} {\bibfnamefont {A.}~\bibnamefont {Okafuji}}, \bibinfo
  {author} {\bibfnamefont {A.~R.}\ \bibnamefont {Marino}}, \bibinfo {author}
  {\bibfnamefont {T.}~\bibnamefont {Berthold}}, \bibinfo {author}
  {\bibfnamefont {G.}~\bibnamefont {Link}}, \bibinfo {author} {\bibfnamefont
  {K.}~\bibnamefont {Hitomi}}, \bibinfo {author} {\bibfnamefont {E.~D.}\
  \bibnamefont {Getzoff}}, \bibinfo {author} {\bibfnamefont {E.}~\bibnamefont
  {Schleicher}}, \ and\ \bibinfo {author} {\bibfnamefont {J.~R.}\ \bibnamefont
  {Norris}, \bibfnamefont {Jr.}},\ }\href@noop {} {\bibfield  {journal}
  {\bibinfo  {journal} {J.\ Phys.\ Chem.\ B}\ }\textbf {\bibinfo {volume}
  {114}},\ \bibinfo {pages} {14745} (\bibinfo {year} {2010})}\BibitemShut
  {NoStop}%
\bibitem [{\citenamefont {Steiner}\ and\ \citenamefont
  {Ulrich}(1989)}]{stei-cr-89-51}%
  \BibitemOpen
  \bibfield  {author} {\bibinfo {author} {\bibfnamefont {U.~E.}\ \bibnamefont
  {Steiner}}\ and\ \bibinfo {author} {\bibfnamefont {T.}~\bibnamefont
  {Ulrich}},\ }\href@noop {} {\bibfield  {journal} {\bibinfo  {journal} {Chem.\
  Rev.}\ }\textbf {\bibinfo {volume} {89}},\ \bibinfo {pages} {51} (\bibinfo
  {year} {1989})}\BibitemShut {NoStop}%
\bibitem [{\citenamefont {Maeda}\ \emph {et~al.}(2011)\citenamefont {Maeda},
  \citenamefont {Wedge}, \citenamefont {Storey}, \citenamefont {Henbest},
  \citenamefont {Liddell}, \citenamefont {Kodis}, \citenamefont {Gust},
  \citenamefont {Hore},\ and\ \citenamefont {Timmel}}]{maed-cc-47-6563}%
  \BibitemOpen
  \bibfield  {author} {\bibinfo {author} {\bibfnamefont {K.}~\bibnamefont
  {Maeda}}, \bibinfo {author} {\bibfnamefont {C.~J.}\ \bibnamefont {Wedge}},
  \bibinfo {author} {\bibfnamefont {J.~G.}\ \bibnamefont {Storey}}, \bibinfo
  {author} {\bibfnamefont {K.~B.}\ \bibnamefont {Henbest}}, \bibinfo {author}
  {\bibfnamefont {P.~A.}\ \bibnamefont {Liddell}}, \bibinfo {author}
  {\bibfnamefont {G.}~\bibnamefont {Kodis}}, \bibinfo {author} {\bibfnamefont
  {D.}~\bibnamefont {Gust}}, \bibinfo {author} {\bibfnamefont {P.~J.}\
  \bibnamefont {Hore}}, \ and\ \bibinfo {author} {\bibfnamefont {C.~R.}\
  \bibnamefont {Timmel}},\ }\href@noop {} {\bibfield  {journal} {\bibinfo
  {journal} {Chem.\ Commun.}\ }\textbf {\bibinfo {volume} {47}},\ \bibinfo
  {pages} {6563} (\bibinfo {year} {2011})}\BibitemShut {NoStop}%
\bibitem [{\citenamefont {de~Groot}\ \emph {et~al.}(1967)\citenamefont
  {de~Groot}, \citenamefont {Hesselmann},\ and\ \citenamefont {van~der
  Waals}}]{groo-mp-12-259}%
  \BibitemOpen
  \bibfield  {author} {\bibinfo {author} {\bibfnamefont {M.~S.}\ \bibnamefont
  {de~Groot}}, \bibinfo {author} {\bibfnamefont {I.~A.~M.}\ \bibnamefont
  {Hesselmann}}, \ and\ \bibinfo {author} {\bibfnamefont {J.~H.}\ \bibnamefont
  {van~der Waals}},\ }\href@noop {} {\bibfield  {journal} {\bibinfo  {journal}
  {Mol.\ Phys.}\ }\textbf {\bibinfo {volume} {12}},\ \bibinfo {pages} {259}
  (\bibinfo {year} {1967})}\BibitemShut {NoStop}%
\bibitem [{\citenamefont {Atkins}\ and\ \citenamefont
  {Evans}(1974)}]{atki-mp-27-1633}%
  \BibitemOpen
  \bibfield  {author} {\bibinfo {author} {\bibfnamefont {P.~W.}\ \bibnamefont
  {Atkins}}\ and\ \bibinfo {author} {\bibfnamefont {G.~T.}\ \bibnamefont
  {Evans}},\ }\href@noop {} {\bibfield  {journal} {\bibinfo  {journal} {Mol.\
  Phys.}\ }\textbf {\bibinfo {volume} {27}},\ \bibinfo {pages} {1633} (\bibinfo
  {year} {1974})}\BibitemShut {NoStop}%
\bibitem [{\citenamefont {Katsuki}\ \emph {et~al.}(2002)\citenamefont
  {Katsuki}, \citenamefont {Kobori}, \citenamefont {Tero-Kubota}, \citenamefont
  {Milikisyants}, \citenamefont {Paul},\ and\ \citenamefont
  {Steiner}}]{kats-mp-100-1245}%
  \BibitemOpen
  \bibfield  {author} {\bibinfo {author} {\bibfnamefont {A.}~\bibnamefont
  {Katsuki}}, \bibinfo {author} {\bibfnamefont {Y.}~\bibnamefont {Kobori}},
  \bibinfo {author} {\bibfnamefont {S.}~\bibnamefont {Tero-Kubota}}, \bibinfo
  {author} {\bibfnamefont {S.}~\bibnamefont {Milikisyants}}, \bibinfo {author}
  {\bibfnamefont {H.}~\bibnamefont {Paul}}, \ and\ \bibinfo {author}
  {\bibfnamefont {U.~E.}\ \bibnamefont {Steiner}},\ }\href@noop {} {\bibfield
  {journal} {\bibinfo  {journal} {Mol.\ Phys.}\ }\textbf {\bibinfo {volume}
  {100}},\ \bibinfo {pages} {1245} (\bibinfo {year} {2002})}\BibitemShut
  {NoStop}%
\bibitem [{\citenamefont {Kothe}\ \emph {et~al.}(2010)\citenamefont {Kothe},
  \citenamefont {Yago}, \citenamefont {Weidner}, \citenamefont {Link},
  \citenamefont {Lukaschek},\ and\ \citenamefont {Lin}}]{koth-jpcb-114-14755}%
  \BibitemOpen
  \bibfield  {author} {\bibinfo {author} {\bibfnamefont {G.}~\bibnamefont
  {Kothe}}, \bibinfo {author} {\bibfnamefont {T.}~\bibnamefont {Yago}},
  \bibinfo {author} {\bibfnamefont {J.-U.}\ \bibnamefont {Weidner}}, \bibinfo
  {author} {\bibfnamefont {G.}~\bibnamefont {Link}}, \bibinfo {author}
  {\bibfnamefont {M.}~\bibnamefont {Lukaschek}}, \ and\ \bibinfo {author}
  {\bibfnamefont {T.-S.}\ \bibnamefont {Lin}},\ }\href@noop {} {\bibfield
  {journal} {\bibinfo  {journal} {J.\ Phys.\ Chem.\ B}\ }\textbf {\bibinfo
  {volume} {114}},\ \bibinfo {pages} {14755} (\bibinfo {year}
  {2010})}\BibitemShut {NoStop}%
\bibitem [{\citenamefont {Timmel}\ \emph {et~al.}(1998)\citenamefont {Timmel},
  \citenamefont {Till}, \citenamefont {Brocklehurst}, \citenamefont
  {McLauchlan},\ and\ \citenamefont {Hore}}]{timm-mp-95-71}%
  \BibitemOpen
  \bibfield  {author} {\bibinfo {author} {\bibfnamefont {C.~R.}\ \bibnamefont
  {Timmel}}, \bibinfo {author} {\bibfnamefont {U.}~\bibnamefont {Till}},
  \bibinfo {author} {\bibfnamefont {B.}~\bibnamefont {Brocklehurst}}, \bibinfo
  {author} {\bibfnamefont {K.~A.}\ \bibnamefont {McLauchlan}}, \ and\ \bibinfo
  {author} {\bibfnamefont {P.~J.}\ \bibnamefont {Hore}},\ }\href@noop {}
  {\bibfield  {journal} {\bibinfo  {journal} {Mol.\ Phys.}\ }\textbf {\bibinfo
  {volume} {95}},\ \bibinfo {pages} {71} (\bibinfo {year} {1998})}\BibitemShut
  {NoStop}%
\bibitem [{\citenamefont {Timmel}\ \emph {et~al.}(2001)\citenamefont {Timmel},
  \citenamefont {Cintolesi}, \citenamefont {Brocklehurst},\ and\ \citenamefont
  {Hore}}]{timm-cpl-334-387}%
  \BibitemOpen
  \bibfield  {author} {\bibinfo {author} {\bibfnamefont {C.~R.}\ \bibnamefont
  {Timmel}}, \bibinfo {author} {\bibfnamefont {F.}~\bibnamefont {Cintolesi}},
  \bibinfo {author} {\bibfnamefont {B.}~\bibnamefont {Brocklehurst}}, \ and\
  \bibinfo {author} {\bibfnamefont {P.~J.}\ \bibnamefont {Hore}},\ }\href@noop
  {} {\bibfield  {journal} {\bibinfo  {journal} {Chem.\ Phys.\ Lett.}\ }\textbf
  {\bibinfo {volume} {334}},\ \bibinfo {pages} {387} (\bibinfo {year}
  {2001})}\BibitemShut {NoStop}%
\bibitem [{\citenamefont {Wootters}(1998)}]{woot-prl-80-2245}%
  \BibitemOpen
  \bibfield  {author} {\bibinfo {author} {\bibfnamefont {W.~K.}\ \bibnamefont
  {Wootters}},\ }\href@noop {} {\bibfield  {journal} {\bibinfo  {journal}
  {Phys.\ Rev.\ Lett.}\ }\textbf {\bibinfo {volume} {80}},\ \bibinfo {pages}
  {2245} (\bibinfo {year} {1998})}\BibitemShut {NoStop}%
\bibitem [{\citenamefont {Katsoprinakis}\ \emph {et~al.}(2010)\citenamefont
  {Katsoprinakis}, \citenamefont {Dellis},\ and\ \citenamefont
  {Kominis}}]{kats-njp-12-085016}%
  \BibitemOpen
  \bibfield  {author} {\bibinfo {author} {\bibfnamefont {G.~E.}\ \bibnamefont
  {Katsoprinakis}}, \bibinfo {author} {\bibfnamefont {A.~T.}\ \bibnamefont
  {Dellis}}, \ and\ \bibinfo {author} {\bibfnamefont {I.~K.}\ \bibnamefont
  {Kominis}},\ }\href@noop {} {\bibfield  {journal} {\bibinfo  {journal} {New
  J.\ Phys.}\ }\textbf {\bibinfo {volume} {12}},\ \bibinfo {pages} {085016}
  (\bibinfo {year} {2010})}\BibitemShut {NoStop}%
\bibitem [{\citenamefont {Langenbacher}\ \emph {et~al.}(2009)\citenamefont
  {Langenbacher}, \citenamefont {Immeln}, \citenamefont {Dick},\ and\
  \citenamefont {Kottke}}]{lang-jacs-131-14274}%
  \BibitemOpen
  \bibfield  {author} {\bibinfo {author} {\bibfnamefont {T.}~\bibnamefont
  {Langenbacher}}, \bibinfo {author} {\bibfnamefont {D.}~\bibnamefont
  {Immeln}}, \bibinfo {author} {\bibfnamefont {B.}~\bibnamefont {Dick}}, \ and\
  \bibinfo {author} {\bibfnamefont {T.}~\bibnamefont {Kottke}},\ }\href@noop {}
  {\bibfield  {journal} {\bibinfo  {journal} {J.\ Am.\ Chem.\ Soc.}\ }\textbf
  {\bibinfo {volume} {131}},\ \bibinfo {pages} {14274} (\bibinfo {year}
  {2009})}\BibitemShut {NoStop}%
\bibitem [{\citenamefont {Weber}\ \emph {et~al.}(2001)\citenamefont {Weber},
  \citenamefont {M{\"o}bius}, \citenamefont {Richter},\ and\ \citenamefont
  {Kay}}]{webe-jacs-123-3790}%
  \BibitemOpen
  \bibfield  {author} {\bibinfo {author} {\bibfnamefont {S.}~\bibnamefont
  {Weber}}, \bibinfo {author} {\bibfnamefont {K.}~\bibnamefont {M{\"o}bius}},
  \bibinfo {author} {\bibfnamefont {G.}~\bibnamefont {Richter}}, \ and\
  \bibinfo {author} {\bibfnamefont {C.~W.~M.}\ \bibnamefont {Kay}},\
  }\href@noop {} {\bibfield  {journal} {\bibinfo  {journal} {J.\ Am.\ Chem.\
  Soc.}\ }\textbf {\bibinfo {volume} {123}},\ \bibinfo {pages} {3790} (\bibinfo
  {year} {2001})}\BibitemShut {NoStop}%
\bibitem [{\citenamefont {Biskup}\ \emph {et~al.}(2011)\citenamefont {Biskup},
  \citenamefont {Hitomi}, \citenamefont {Getzoff}, \citenamefont {Krapf},
  \citenamefont {Koslowski}, \citenamefont {Schleicher},\ and\ \citenamefont
  {Weber}}]{bisk-acie-50-12647}%
  \BibitemOpen
  \bibfield  {author} {\bibinfo {author} {\bibfnamefont {T.}~\bibnamefont
  {Biskup}}, \bibinfo {author} {\bibfnamefont {K.}~\bibnamefont {Hitomi}},
  \bibinfo {author} {\bibfnamefont {E.~D.}\ \bibnamefont {Getzoff}}, \bibinfo
  {author} {\bibfnamefont {S.}~\bibnamefont {Krapf}}, \bibinfo {author}
  {\bibfnamefont {T.}~\bibnamefont {Koslowski}}, \bibinfo {author}
  {\bibfnamefont {E.}~\bibnamefont {Schleicher}}, \ and\ \bibinfo {author}
  {\bibfnamefont {S.}~\bibnamefont {Weber}},\ }\href@noop {} {\bibfield
  {journal} {\bibinfo  {journal} {Angew.\ Chem.\ Int.\ Ed.}\ }\textbf {\bibinfo
  {volume} {50}},\ \bibinfo {pages} {12647} (\bibinfo {year}
  {2011})}\BibitemShut {NoStop}%
\bibitem [{\citenamefont {Eisenreich}\ \emph {et~al.}(2008)\citenamefont
  {Eisenreich}, \citenamefont {Joshi}, \citenamefont {Weber}, \citenamefont
  {Bacher},\ and\ \citenamefont {Fischer}}]{eise-jacs-130-13544}%
  \BibitemOpen
  \bibfield  {author} {\bibinfo {author} {\bibfnamefont {W.}~\bibnamefont
  {Eisenreich}}, \bibinfo {author} {\bibfnamefont {M.}~\bibnamefont {Joshi}},
  \bibinfo {author} {\bibfnamefont {S.}~\bibnamefont {Weber}}, \bibinfo
  {author} {\bibfnamefont {A.}~\bibnamefont {Bacher}}, \ and\ \bibinfo {author}
  {\bibfnamefont {M.}~\bibnamefont {Fischer}},\ }\href@noop {} {\bibfield
  {journal} {\bibinfo  {journal} {J.\ Am.\ Chem.\ Soc.}\ }\textbf {\bibinfo
  {volume} {130}},\ \bibinfo {pages} {13544} (\bibinfo {year}
  {2008})}\BibitemShut {NoStop}%
\bibitem [{\citenamefont {Thamarath}\ \emph {et~al.}(2010)\citenamefont
  {Thamarath}, \citenamefont {Heberle}, \citenamefont {Hore}, \citenamefont
  {Kottke},\ and\ \citenamefont {Matysik}}]{tham-jacs-132-15542}%
  \BibitemOpen
  \bibfield  {author} {\bibinfo {author} {\bibfnamefont {S.~S.}\ \bibnamefont
  {Thamarath}}, \bibinfo {author} {\bibfnamefont {J.}~\bibnamefont {Heberle}},
  \bibinfo {author} {\bibfnamefont {P.~J.}\ \bibnamefont {Hore}}, \bibinfo
  {author} {\bibfnamefont {T.}~\bibnamefont {Kottke}}, \ and\ \bibinfo {author}
  {\bibfnamefont {J.}~\bibnamefont {Matysik}},\ }\href@noop {} {\bibfield
  {journal} {\bibinfo  {journal} {J.\ Am.\ Chem.\ Soc.}\ }\textbf {\bibinfo
  {volume} {132}},\ \bibinfo {pages} {15542} (\bibinfo {year}
  {2010})}\BibitemShut {NoStop}%
\bibitem [{\citenamefont {Kowalczyk}\ \emph {et~al.}(2004)\citenamefont
  {Kowalczyk}, \citenamefont {Schleicher}, \citenamefont {Bittl},\ and\
  \citenamefont {Weber}}]{kowa-jacs-126-11393}%
  \BibitemOpen
  \bibfield  {author} {\bibinfo {author} {\bibfnamefont {R.~M.}\ \bibnamefont
  {Kowalczyk}}, \bibinfo {author} {\bibfnamefont {E.}~\bibnamefont
  {Schleicher}}, \bibinfo {author} {\bibfnamefont {R.}~\bibnamefont {Bittl}}, \
  and\ \bibinfo {author} {\bibfnamefont {S.}~\bibnamefont {Weber}},\
  }\href@noop {} {\bibfield  {journal} {\bibinfo  {journal} {J.\ Am.\ Chem.\
  Soc.}\ }\textbf {\bibinfo {volume} {126}},\ \bibinfo {pages} {11393}
  (\bibinfo {year} {2004})}\BibitemShut {NoStop}%
\bibitem [{\citenamefont {Hore}(1980)}]{hore-cpl-69-563}%
  \BibitemOpen
  \bibfield  {author} {\bibinfo {author} {\bibfnamefont {P.~J.}\ \bibnamefont
  {Hore}},\ }\href@noop {} {\bibfield  {journal} {\bibinfo  {journal} {Chem.\
  Phys.\ Lett.}\ }\textbf {\bibinfo {volume} {69}},\ \bibinfo {pages} {563}
  (\bibinfo {year} {1980})}\BibitemShut {NoStop}%
\bibitem [{\citenamefont {Buckley}\ \emph {et~al.}(1987)\citenamefont
  {Buckley}, \citenamefont {Hunter}, \citenamefont {Hore},\ and\ \citenamefont
  {McLauchlan}}]{buck-cpl-135-307}%
  \BibitemOpen
  \bibfield  {author} {\bibinfo {author} {\bibfnamefont {C.~D.}\ \bibnamefont
  {Buckley}}, \bibinfo {author} {\bibfnamefont {D.~A.}\ \bibnamefont {Hunter}},
  \bibinfo {author} {\bibfnamefont {P.~J.}\ \bibnamefont {Hore}}, \ and\
  \bibinfo {author} {\bibfnamefont {K.~A.}\ \bibnamefont {McLauchlan}},\
  }\href@noop {} {\bibfield  {journal} {\bibinfo  {journal} {Chem.\ Phys.\
  Lett.}\ }\textbf {\bibinfo {volume} {135}},\ \bibinfo {pages} {307} (\bibinfo
  {year} {1987})}\BibitemShut {NoStop}%
\bibitem [{\citenamefont {Hore}\ \emph {et~al.}(1987)\citenamefont {Hore},
  \citenamefont {Hunter}, \citenamefont {McKie},\ and\ \citenamefont
  {Hoff}}]{hore-cpl-137-495}%
  \BibitemOpen
  \bibfield  {author} {\bibinfo {author} {\bibfnamefont {P.~J.}\ \bibnamefont
  {Hore}}, \bibinfo {author} {\bibfnamefont {D.~A.}\ \bibnamefont {Hunter}},
  \bibinfo {author} {\bibfnamefont {C.~D.}\ \bibnamefont {McKie}}, \ and\
  \bibinfo {author} {\bibfnamefont {A.~J.}\ \bibnamefont {Hoff}},\ }\href@noop
  {} {\bibfield  {journal} {\bibinfo  {journal} {Chem.\ Phys.\ Lett.}\ }\textbf
  {\bibinfo {volume} {137}},\ \bibinfo {pages} {495} (\bibinfo {year}
  {1987})}\BibitemShut {NoStop}%
\end{thebibliography}%

\end{document}